\documentclass[dvipsnames,twocolumn]{aastex63}
\PassOptionsToPackage{usenames,dvipsnames}{xcolor}
\usepackage{tabularx}
\usepackage{amsmath}  
\usepackage{amssymb}  
\usepackage[T1]{fontenc}
\usepackage{xspace}
\usepackage{booktabs}
\usepackage{threeparttable}

\setcitestyle{square,numbers}

\gdef\arcsec{$^{\prime\prime}$}

\def\SNABC{AT\,2016jka\xspace}

\def\lenstool{{\tt LENSTOOL}\xspace}

\submitjournal{a peer-reviewed journal}
\accepted{June 8, 2021}
\shorttitle{SN Requiem}
\shortauthors{Rodney et al.}    

\begin{document}

\title{A Gravitationally Lensed Supernova with an Observable Two-Decade Time Delay}

\correspondingauthor{S. Rodney \& G. Brammer}
\email{srodney@sc.edu}; \email{gabriel.brammer@nbi.ku.dk}

\author{Steven A. Rodney}
 
\affil{Department of Physics \& Astronomy, University of South Carolina,Columbia, SC 29208, USA}
\author{Gabriel B. Brammer}

\affil{Cosmic Dawn Center (DAWN), Copenhagen Denmark}
\affil{Niels Bohr Institute, University of Copenhagen, Jagtvej 128, Copenhagen, Denmark}
\author{Justin D.\,R.~Pierel} 

\affil{Department of Physics \& Astronomy, University of South Carolina,Columbia, SC 29208, USA}
\author{Johan Richard}

\affil{Univ Lyon, Univ Lyon1, Ens de Lyon, CNRS, Centre de Recherche Astrophysique de Lyon}
\author{Sune Toft}

\affil{Cosmic Dawn Center (DAWN), Copenhagen Denmark}
\affil{Niels Bohr Institute, University of Copenhagen, Jagtvej 128, Copenhagen, Denmark}
\author{Kyle F. O'Connor}

\affil{Department of Physics \& Astronomy, University of South Carolina,Columbia, SC 29208, USA}

\author{Mohammad Akhshik} 

\affil{Department of Physics, University of Connecticut,Storrs, CT 06269, USA}
\author{Katherine E. Whitaker} 
\affil{Department of Astronomy, University of Massachusetts,Amherst, MA 01003, USA}
\affil{Cosmic Dawn Center (DAWN), Copenhagen Denmark}
\begin{abstract}
  When the light from a distant object passes very near to a foreground galaxy or cluster, gravitational lensing can cause it to appear as multiple images on the sky \cite{einstein_lens_1936}.  
  If the source is variable, it can be used to constrain the cosmic expansion rate \cite{refsdal_possibility_1964} and dark energy models \cite{holz_seeing_2001}.  
  Achieving these cosmological goals requires many lensed transients with precise time delay measurements \cite{treu_time_2016}. 
  Lensed SN are attractive for this purpose because they have relatively simple photometric behavior, with well-understood light curve shapes and colours---in contrast to the stochastic variation of quasars.
  Here we report the discovery of a multiply-imaged supernova, \SNABC (``SN Requiem''). It appeared in an evolved galaxy at $z=1.95$, gravitationally lensed by a foreground galaxy cluster \cite{brammer_discovery_2021}. 
  It is likely a Type Ia supernova---the explosion of a low-mass stellar remnant, whose light curve can be used to measure cosmic distances.  
  In archival Hubble Space Telescope imaging, three lensed images of the supernova are detected with relative time delays of $<$200 days.  
  We predict a fourth image will appear close to the cluster core in the year 2037$\pm$2. 
  Observation of the fourth image could provide a time delay precision of $\approx 7$ days, $<1\%$ of the extraordinary 20 year baseline. 
  The SN classification and the predicted reappearance time could be improved with further lens modelling and a comprehensive analysis of systematic uncertainties.
  
\end{abstract}

\section*{}

\begin{figure*}
    \centering
    \includegraphics[width=\textwidth]{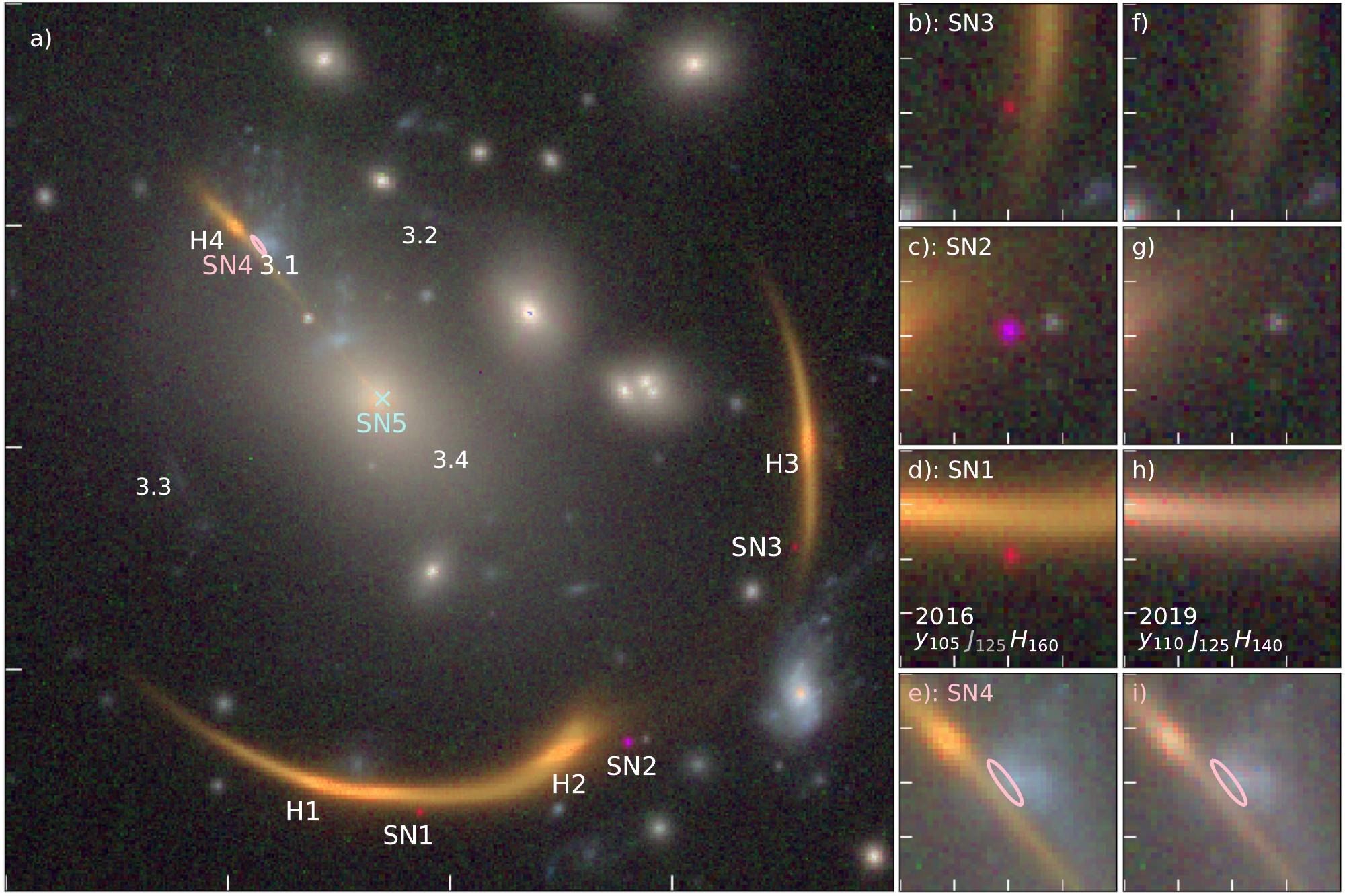}
    \caption{\textbf{Overview of the MACS J0138.0-2155 cluster field and the \SNABC discovery.}  The wide-field view in $a)$ is 40\arcsec\ on a side with ticks indicating 10\arcsec\ intervals. 
    Labels indicate the locations of the observed\SNABC images (SN1--SN3), the expected future images (SN4--SN5), and the multiply-imaged host galaxy (H1--H4). 
    Labels 3.1--3.4 indicate the locations of a separate  multiply-imaged [O\textsc{ii}]-emitter at $z=0.77$ used to help constrain the lensing potential. 
    Panels \emph{b}--\emph{i} show 4\arcsec\ cutouts around the lensed SN images with 1\arcsec\ ticks.   
    Panels \emph{bcde} show the imaging from 2016 July where the SN was visible and \emph{fghi} show the later imaging from 2019 July where the the SN has faded away.  
    Panels \emph{aei} indicate the location of the fourth image predicted to appear in $\approx$2037, with an ellipse demarcating the 68\% probability region predicted from the \lenstool model.  
    Panel \emph{a} shows the expected location of the final and highly demagnified fifth image (``SN5''). 
    The three-colour images are generated from the WFC3/IR filters with R=F105W or F110W, G=F125W and R=F160W or F140W (as indicated in panels \emph{dh}).    
    All panels use the late-epoch F125W imaging for the green channel; nevertheless, it is immediately clear that the SN2 image is substantially bluer than the other two, which helps to constrain the relative age of each SN image and the transient classification as a likely Type Ia supernova explosion. }
    \label{fig:layout}
\end{figure*}

We discovered \SNABC 
using data from the Hubble Space Telescope (HST) program {\it REsolved QUIEscent Magnified Galaxies} (REQUIEM, HST-GO-15663, PI:Akhshik) \cite{akhshik_recent_2021} (see a summary of observations in Supplementary Table~1). The REQUIEM project targets massive galaxies with low specific star formation rates that have been magnified by strong gravitational lensing. 
The brightest and most spectacular galaxy targeted by REQUIEM
is MRG-M0138, a massive red galaxy (MRG) at $z=1.95$ \cite[ref][]{newman_resolving_2018} behind the galaxy cluster MACS J0138.0-2155 \cite[ref][]{ebeling_macs_2001}.
MRG-M0138 is quadruply lensed by a foreground galaxy cluster at $z=0.338$.  
During analysis of observations obtained 13--14 July 2019
we discovered three point 
sources that were present in archival HST images from 18--19 July 2016, part of the program (HST-GO-14496; PI:Newman)
that first confirmed the MRG-M0138 galaxy as a strongly-lensed object 
(Methods: Observations). 
Each point source is within 5 arcseconds of one of the four MRG-M0138 images.  None of the
three point sources are present in the REQUIEM HST data in 2019 (Fig. \ref{fig:layout}). 
We infer that these are multiple images of a single astrophysical transient in MRG-M0138, most likely a SN. 

\begin{table*}[tb]
    \caption{\label{tab:time_delays} SN lensing observables.} 
    \centering
\begin{threeparttable}
    \begin{tabular}{cccllcc}
    Image     & R.A.\tnote{a} & Dec.\tnote{a} & 
    Age\tnote{b} &  $\Delta t_{\rm obs}$\tnote{c} 
    & $\Delta t_{\rm pred}$\tnote{c} 
    & $\mu_{\rm pred}$ \\
    & [hh:mm:ss] & [dd:mm:ss] &[days] & [days] & [days] &  \\
\hline
SN1 & 01:38:03.63     & $-$21:55:50.38        & $92^{+21}_{-19}$  & ---- & --- & 5$\pm$1 \\[4pt]
SN2 & 01:38:02.96     & $-$21:55:47.26        & $-24^{+16}_{-7}$  & $114^{+28}_{-31}$ & $82\pm62$  & $7\pm3$ \\[4pt]
SN3 & 01:38:02.42     & $-$21:55:38.47        & $107^{+26}_{-21}$ & $-17^{+19}_{-16}$ & $-19\pm34$ & $3.9\pm0.5$  \\[4pt]
SN4 & 01:38:04.15$\pm$0.36 & $-$21:55:24.73$\pm$0.43 &  --- & --- & $7742\pm540$\tnote{d} & $0.4\pm0.2$ \\[4pt]
SN5 & 01:38:03.77$\pm$0.1  & $-$21:55:31.74$\pm$0.1                    & --- & --- & $9463\pm770$\tnote{d} & $<0.01$\\
\hline 
\end{tabular}

\begin{tablenotes}    
    \item[a]{Coordinates are given in the J2000 reference frame, as measured for images 1--3 and as predicted for images 4 and 5.  Uncertainties in the predicted position of images 4 and 5 are in arcseconds.}
    \item[b]{The measured age of each SN image is the number of days relative to peak brightness, in observer-frame days. A negative value means the image was observed prior to peak brightness.  }
    \item[c]{Time delays are reported relative to image SN1.}
    \item[d]{The combination of the models for the light curve evolution and the lens time delays implies a date of $2037\pm2$ for the peak brightness of Image 4 and 2042$\pm$4 for Image 5 (which will likely be undetectable).}
\end{tablenotes}
\end{threeparttable}
\end{table*}

To construct a lens model for the MACS J0138.0-2155 cluster we use the \lenstool software \cite{jullo_bayesian_2007,kneib_lenstool_2011} (Methods: Lens Modelling). 
To avoid unintended bias, we kept the lens model development completely separate from the analysis of the SN.  Only upon  completion of both were the results combined for the analysis described here.  The input model constraints are the positions and redshifts of the MRG-M0138 galaxy at $z=1.95$ (both the galaxy's centroid position and the SN location in each image) as well as a multiply-imaged background 
galaxy at $z=0.766$, both having secure spectroscopic redshifts (Supplementary Table 3). 
We model the mass distribution in the cluster core as the combination of a cluster-scale and galaxy-scale potentials (Supplementary Table 4; Extended Data Figure 1).
From this model we derive estimates for the lensing magnification and time delay of each of the SN images, including two predicted future images (Table \ref{tab:time_delays}).
The lens model predicts that the SN should appear in the fourth MRG-M0138 image in the year 2037$\pm$2, demagnified with $\mu=0.4\pm0.2$. 
A fifth image will also appear at a still later date, located near the center of the cluster and much more significantly demagnified, so it will not be easily observable. 
We anticipate that future lens modelling of the cluster will improve on these predictions primarily by exploring a wider range of mass models and incorporating more observational constraints (Supplementary Note: Future Work).   
For example, our analysis adopted only a single form for the density profiles, and did not incorporate constraints from stellar kinematics or pixel-level surface brightness data from the multiply-imaged systems. Although our \lenstool model does a good job of reproducing the morphology of the host galaxy images H1-H3, it does not reproduce image H4 as well (Supplementary Note: Host Image Morphology Comparison). 
 
If we can estimate the age of each SN image then we can derive direct observational constraints on the relative lensing time delays. 
For this goal, it is helpful to have a firm determination of the transient’s class. 
Expected time delays and magnifications from the lens model exclude any of the various rapidly evolving and low-luminosity stellar transient classes, strongly suggesting that it is a SN. 
The first-order SN distinction remaining is between a Type Ia SN---the explosion of a white dwarf star in a binary system---and a core collapse SN (CCSN)---the end-point of a star with mass $>10 M_{\odot}$. 
The properties of the host galaxy can inform this classification because CCSN are limited to galaxies with young stellar populations. 
Limits on the specific star formation rate and age for this host, MRG-M0138, show it to be a massive but very quiescent and evolved galaxy, 
unlikely to retain any significant population of high-mass stars \cite{newman_resolving_2018}. 
Based on observed properties of the host galaxy alone, we find a 62-75\% probability that \SNABC is a Type Ia SN (Methods: Classification). 
\begin{figure*}
  \centering
    \includegraphics[width=0.75\textwidth]{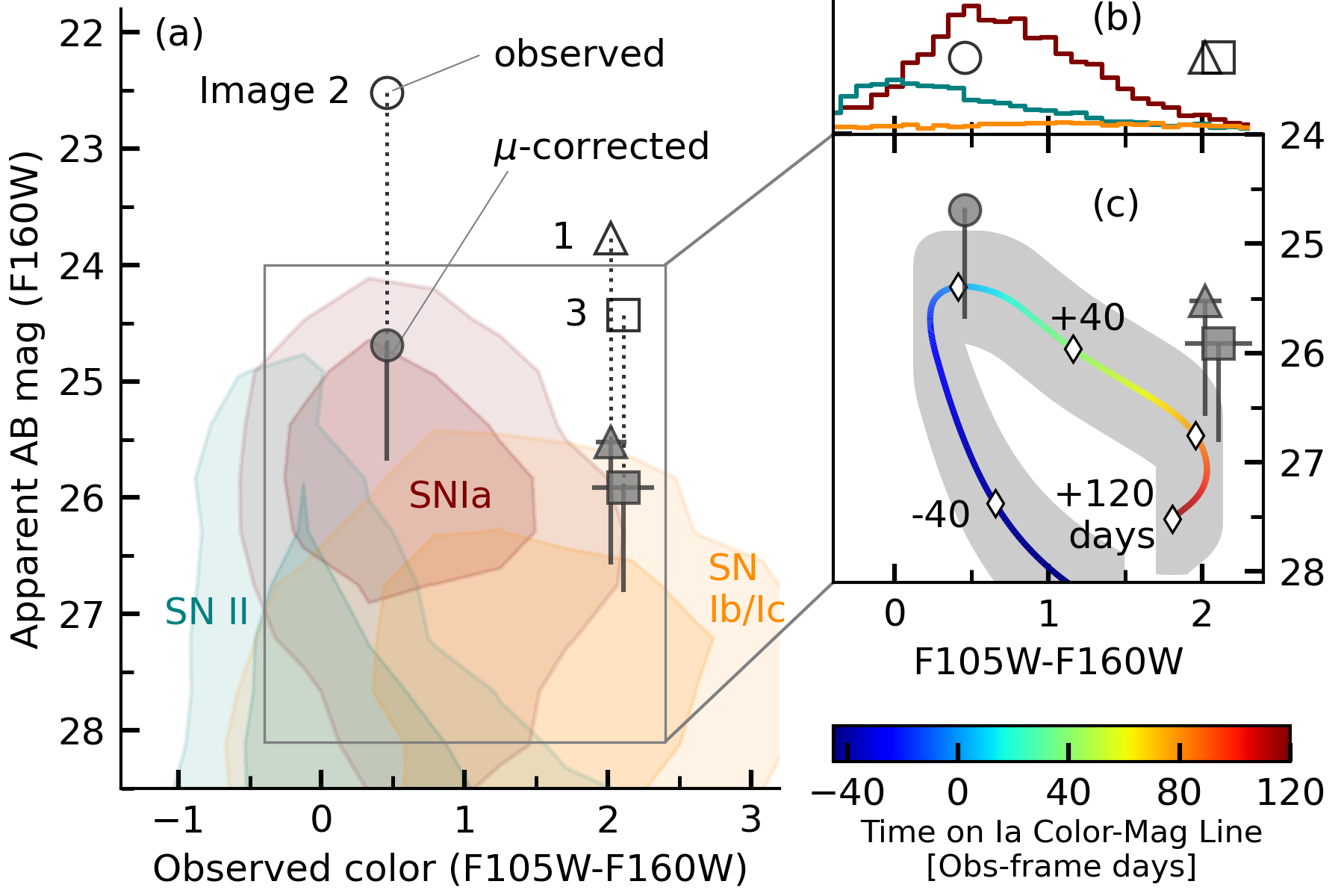}
    \caption{ \textbf{Classification information for \SNABC based on its position in colour-magnitude space.} (a) 
    The observed photometry for the three SN images is shown as open markers. Vertical dotted lines show the lens magnification corrections using our preferred lens model, with a filled grey marker indicating the corrected (demagnified) magnitude. Error bars indicate the observational and systematic uncertainty, including the range of alternative magnification corrections encompassed by lens model variations.
    Contours show the population distributions for normal SNe of Type Ia (red), Type Ib/Ic (gold), and Type II (green), drawn with two contour levels enclosing 68\%  and 95\%  of each SN population.  Each SN sub-class was simulated at $z=1.95$, and samples from their expected light curves were drawn uniformly in time.  
    (b) Marginalized distributions along the colour dimension for the three SN sub-classes (using the same colour scheme).  The simulated populations have been scaled according to the expected explosion rates in the SN host galaxy, based on its stellar population properties, so the y-axis scaling is effectively a relative probability. Open markers show the observed colours again, at arbitrary y position. 
    (c) Zoomed-in view of the colour-mag space marked by the grey box in panel \emph{a}.
    The evolution of a typical Type Ia SN at $z=1.95$  is shown by a coloured line, with the line colour indicating SN age in observer-frame days relative to peak brightness.  White diamonds correspond to the times labeled on the colourbar below. Grey shading shows the typical range of luminosities and colours observed for the Type Ia SN population in the nearby universe. 
    Although the magnification-corrected data are brighter than expected for most Type Ia SN, they are consistent both with the overall Type Ia SN population, and with the Type Ia SN colour-mag vs time curve.  
    \label{fig:class}
    }
\end{figure*}

\begin{figure*}[htbp]
    \centering
    \includegraphics[width=0.49\textwidth]{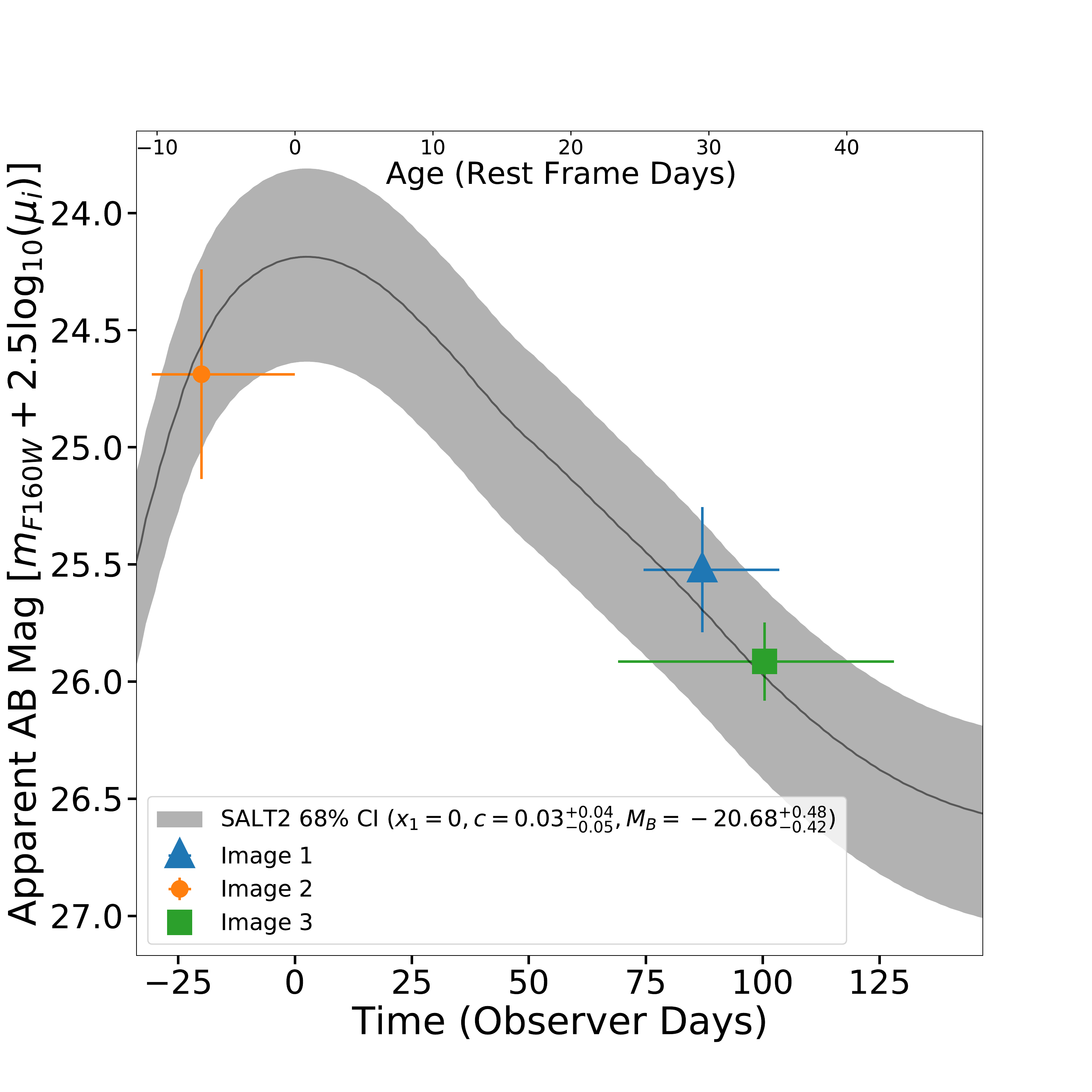}
    \includegraphics[width=0.49\textwidth]{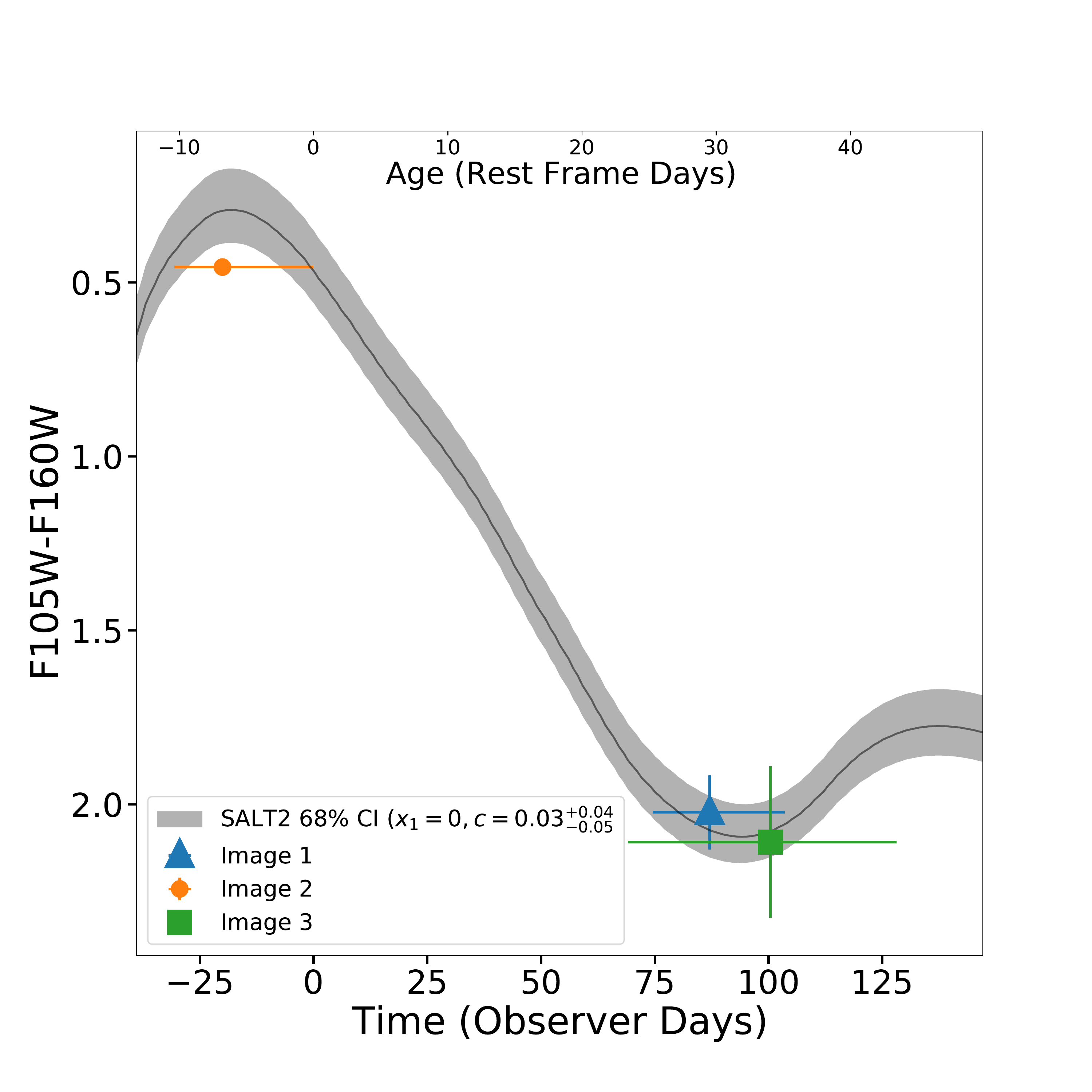}
    \caption{\label{fig:full_lightcurve} \textbf{The reconstructed light curve and colour curve for \SNABC.} Panel (a) shows the light curve and panel (b) shows the colour curve, both incorporating lens model corrections for the time delay and magnification. 
    The grey shaded region covers the 68\% confidence interval of the best-fit Type Ia model (using the SALT2 model \cite{guy_salt2:_2007}), with the median model shown as a solid line. Observed photometric data are shown as coloured markers. The x-dimension error bars on each data point represent the lens model time delay uncertainties (Table 1). The y-dimension uncertainties in both panels incorporate 
    the photometric uncertainty.  
    For the light curve (a), the y error bars also include uncertainty in the lens model magnification (Table 1).  The colour measurements (y values, panel b) do not require lens model correction.  Parameters for the best-fit SALT2 models in grey are shown in the legend, and were obtained using the joint posterior of the colour curve and light curve methods (described in Methods:Time Delay Estimation; shown in Extended Data Fig.~7).}
\end{figure*}

Adopting the lens model magnifications for the three observed SN images (Table \ref{tab:time_delays}) we can locate each SN image in colour-magnitude space (Fig. \ref{fig:class}). 
After magnification correction, all three images are still brighter than expected for a Type Ia SN, which may indicate that a lens-modelling degeneracy is at play. 
Nevertheless, the magnification-corrected \SNABC data are more consistent with the Type Ia population than any CC SN sub-class (Fig.~\ref{fig:class}a,b and Extended Data Fig. 2). \SNABC also demonstrates the expected evolution of a Type Ia SN colour and brightness over $\sim 100$ days (Fig.~\ref{fig:class}c).
By also including the model-predicted time delays, we can treat the three SN images effectively as three points on a common SN light curve, and we find p(Ia)=94\% (Methods: Classification). 
This composite light curve is shown in Figure~\ref{fig:full_lightcurve}, with the best-fitting Type Ia SN model.  Extended Data Fig. 3 also shows a random draw of light curves from the Monte Carlo sampling for all three major SN sub-classes.
For the remainder of this analysis, we proceed under the assumption that \SNABC is indeed a Type Ia SN.  An improved classification could be achieved with spectroscopy and multi-band photometry upon arrival of the fourth image.  
In that case the analysis that follows here could be revised to achieve similar results with a different underlying SN model.

The colour of a Type Ia SN evolves substantially over its lifetime as the photosphere expands and cools, revealing different layers of the expanding shell and driving episodes of recombination \cite{kasen_type_2007}. 
Since the phenomenon of gravitational lensing in general is achromatic, this colour evolution makes it possible to derive an age constraint that is largely independent of the lens model (Methods: Colour Curve Age Constraints; Extended Data Fig. 4 and 5). 
Combining this information with magnification constraints from the lens modelling helps break parameter degeneracies, yielding measured delays in Table \ref{tab:time_delays} (Methods: Light Curve + Lens Model Age Constraints; Extended Data Fig. 6 and 7).

Using these measured time delays, we created a reconstructed form of the intrinsic light curve and colour curve of \SNABC, shown in Fig. \ref{fig:full_lightcurve}.
Remarkably, the ages of image 1 and 3 are constrained to better than $\pm$20 days, despite having only a single epoch of photometric data.    These uncertainties may be 
further reduced when the future fourth image is observed with high-precision, multi-epoch photometry.  Such a light curve will pin down the intrinsic SN light curve parameters that are shared by all images, and break remaining parameter degeneracies. Improvements to the lens modelling will also be essential, to better estimate and minimize systematic biases that may arise from the necessary magnification and time delay corrections.

\SNABC and other lensed SN like it could eventually contribute to mapping the cosmic expansion history and measuring the effects of dark energy, which appears to be driving an accelerating cosmic expansion rate 
\cite{riess_observational_1998,perlmutter_measurements_1999}. 
Recent investigations into the expansion rate of the universe (the Hubble-LeMa\^itre constant; $H_0$) have found that measurements from the local universe are significantly different from the value inferred from measurements of the cosmic microwave background radiation \cite{Riess_large_2019,aghanim_planck_2018}.  
The community is actively attempting to resolve this ``$H_0$ crisis'' by mitigating systematic uncertainties or discovering new physics from the early universe \cite{verde_tensions_2019}. 
Either resolution will require multiple independent cosmological probes.
In recent years, lensed quasar time delays have provided a valuable independent tool for this, with seven high-precision measurements to date
\cite{birrer_tdcosmo_2020}.
As the sample of time delay lenses grows to several dozen (including lensed SNe like \SNABC), it is expected to deliver a measurement of $H_0$ with 1\% precision \cite{birrer_tdcosmo_2021}.

Looking beyond the $H_0$ crisis, determining the nature of dark energy and how it may evolve over time is a primary goal for the large-scale cosmology experiments of the 2020s \cite{spergel_wide_2015,Ivezic_lsst_2019}.
A future sample with $\sim100$ well-measured lensing time delays would be a competitive tool for dark energy studies \cite{treu_time_2016,birrer_tdcosmo_2021}. 
Events like \SNABC could be an important part of this time-delay cosmology sample, but to date there have been only two lensed SN observed with multiple images.  
The first, {\it SN Refsdal}, was a peculiar Type II SN whose image with the longest delay was missed \cite{kelly_multiple_2015,kelly_deja_2016}. 
The second, {\it SN 2016geu}, was a Type Ia SN with short delays that make high precision time delay measurements impossible \cite{goobar_iptf16geu:_2017,dhawan_magnification_2019}. 
An earlier discovery of an unusually luminous SN was also shown to be a strongly-lensed Type Ia SN \cite{quimby_detection_2014}, though the multiple images were not resolved.  
This makes \SNABC just the third discovery of a lensed SN resolved into multiple images.

Future large-scale surveys such as the Vera C. Rubin Observatory 
and the Nancy Grace Roman Space Telescope will observe dozens to hundreds of lensed SN over their mission lifetimes. 
The vast majority of these will be lensed by galaxy-scale deflectors  \cite{goldstein_rates_2019, pierel_projected_2020}  and thus will have significantly shorter delays (of order 10--100 days). 
Since it is the {\it fractional} time delay uncertainty that 
propagates through to any time delay distance measurement, the extraordinarily long time delays of cluster-lensed SNe like \SNABC can deliver significantly better time delay precision, with comparable observational cost (Supplementary Note: Future Discoveries).  
In fact, the long time baseline of lensed SN like \SNABC effectively insures that their cosmological precision is not limited by time delay measurement uncertainty. 
Cluster-scale lenses are more complex than galaxy-scale lenses, but they generally have ``independent'' measurements of the magnification from several multiply-imaged systems in the same field. 
Modelling cluster lenses is very different from galaxy lenses, so objects like \SNABC can provide a valuable check on systematics for the larger sample of transients used in time-delay cosmology.

The first multiply-imaged SN discovery, SN Refsdal, has shown that time delay  cosmography with a cluster-lensed SN is viable \cite{rodney_sn_2016,kelly_deja_2016}. 
However, the observing campaign for  SN Refsdal was extraordinary, deploying more than 75 HST orbits over three years. 
Significant  observational investment has also been required for high-precision time delay measurement of lensed quasars, such as decade-long programs 
\cite{millon_cosmograil_2020} or daily monitoring for high-cadence light curves \cite{millon_tdcosmo_2020}.  
In the case of \SNABC  it will be possible to achieve similar time delay 
precision over the 20-year baseline with just a single imaging epoch as the anchor point. 
A sample of \SNABC-like events could be developed with regular monitoring of cluster-scale lenses, partnered with modest follow-up to characterize 
any lensed SN discovered (Supplementary Note: Future Work). 

HST observations enabled us to find this SN.  We anticipate that HST may be de-orbited and make its final plummet to Earth around the time of the reappearance of \SNABC, so we coin the name {\it SN  Requiem} as an ode to the vast new discovery space that HST continues to unveil.

\clearpage
\bibliographystyle{naturemag}

\section*{Acknowledgments}

The authors wish to thank the anonymous referees for constructive feedback that has substantially improved the presentation of this discovery.  Thanks also to P. Kelly and L. Moustakas for helpful commentary on earlier drafts of this work.
{\bf Data:} Based on observations made with the NASA/ESA Hubble Space Telescope, obtained from the data archive at the Space Telescope Science Institute, and on observations collected at the European Organisation for Astronomical Research in the Southern Hemisphere under ESO programme 0103.A-0777(A).  
{\bf Funding:} Support for this work was provided by NASA through grant numbers HST-GO-14622 (K.E.W.), HST-AR-15050 (J.D.R.P.), HST-GO-15663 (M.A.), and HST-GO-16264 (S.R.) from the Space Telescope Science Institute, which is operated by the Association of Universities for Research in Astronomy, Inc. under NASA contract NAS 5-26555.  The Cosmic Dawn Center of Excellence is funded by the Danish National Research Foundation under grant No. 140. Support was provided by NASA Headquarters under the NASA Future Investigators in Earth and Space Science and Technology (FINESST) awards 80NSSC19K1414 (M.A.) and 80NSSC19K1418 (J.D.R.P.).  K.E.W.\ wishes to acknowledge funding from the Alfred P. Sloan Foundation. 

{\bf Author Contributions:} 
Conceptualization, S.A.R., G.B., and S.T.; Methodology, S.A.R., J.D.R.P., J.R., and K.F.O.; Investigation, G.B., J.R., S.T., M.A., and K.E.W.; Writing - Original Draft, G.B.\ and S.T.; Writing - Review \& Editing, S.A.R., G.B., J.D.R.P., J.R., S.T., K.F.O., M.A., and K.E.W.; Visualization, S.A.R., G.B., J.D.R.P., J.R., K.F.O.; Supervision, S.A.R., G.B., S.T., and K.E.W.; Funding Acquisition, M.A., K.E.W., J.D.R.P., \& S.R.

{\bf Competing Interests:} All authors declare no competing interests.

\clearpage

\renewcommand{\figurename}{Extended Data Figure} 
\setcounter{figure}{0}

\begin{figure}
    \centering
    \includegraphics[width=0.9\textwidth]{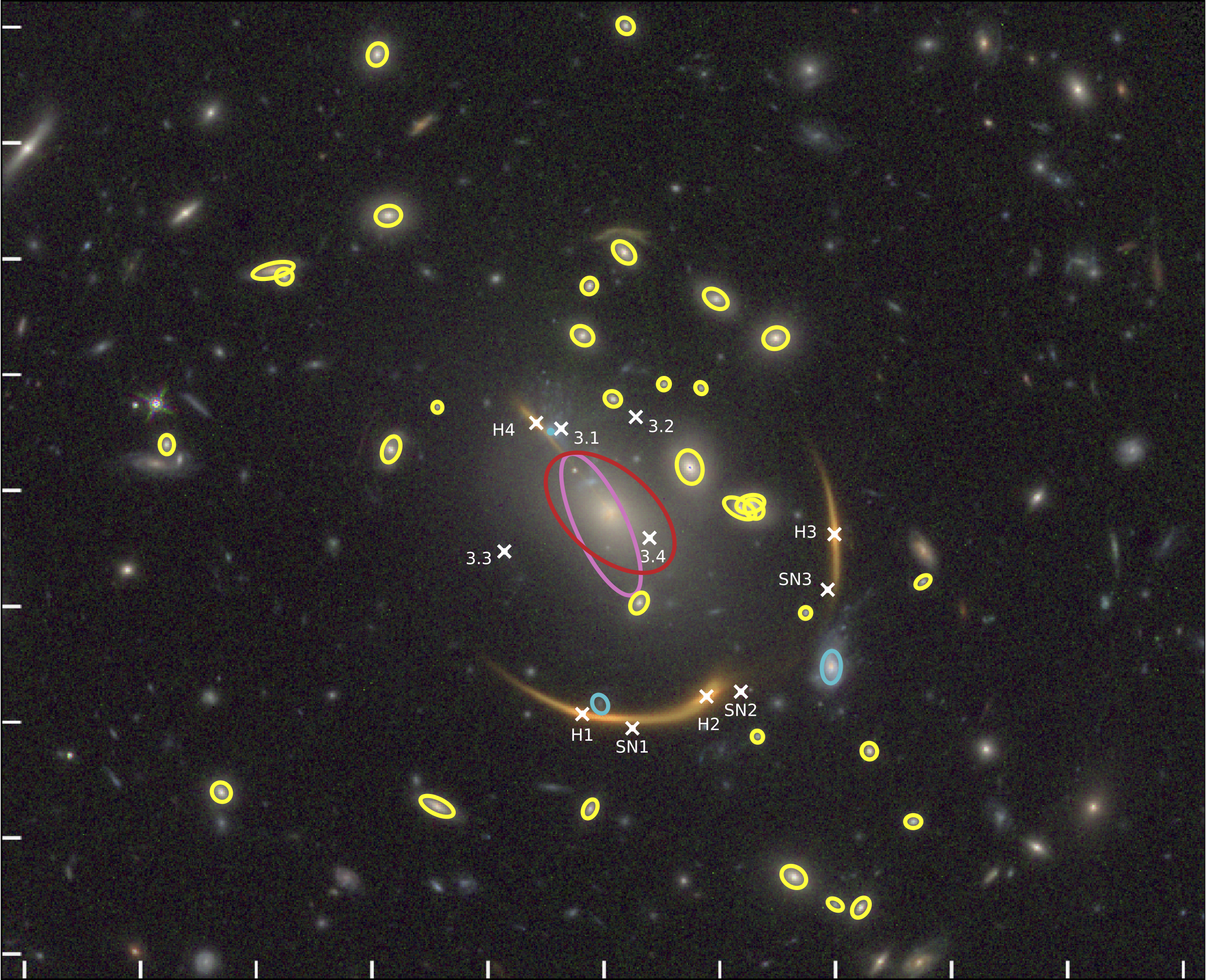}
    \caption{{\bf Elements of the MRG0138 cluster lens model.} The model comprises 37 potentials in total: the BCG (red), 32 cluster members (yellow), three perturbers (cyan), and the main cluster potential (pink).  Labeled × symbols indicate the positions of the SN, host, and one additional multiply-imaged galaxy with a secure redshift used as model constraints (Supplemental Table 3).  The filters used to generate the color image are as in Fig. 1, and tick marks are separated by 10 arcsec.}
    \label{fig:my_label}
\end{figure}

\begin{figure}
    \centering
    \includegraphics[width=0.8\textwidth]{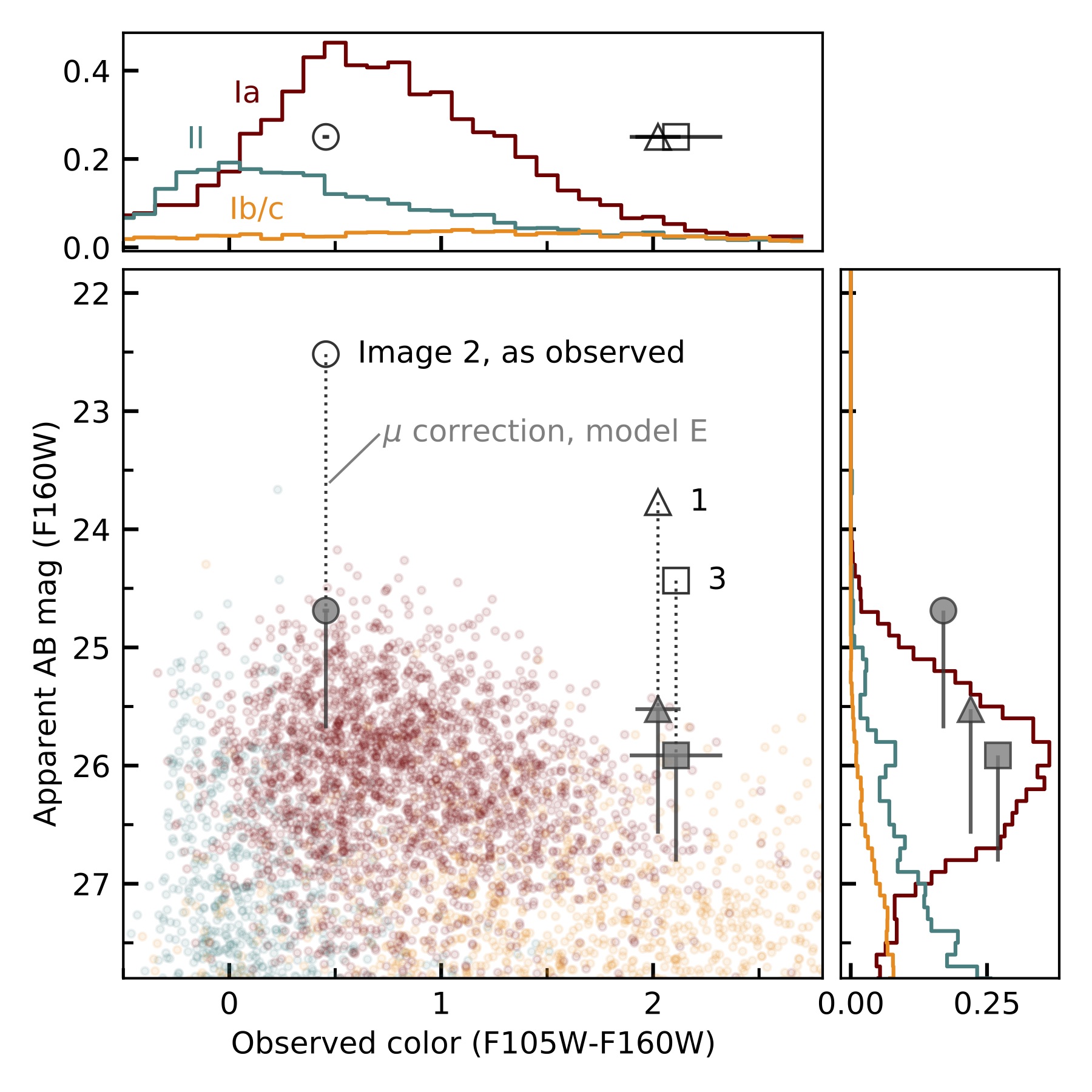}
    \caption{{\bf The position of AT 2016jka in color-magnitude space.}
        Colored points show simulated photometry for normal SNe of Type Ia (red), TypeIb/Ic (gold), and Type II (green), with 10,000 simulated SN in each sub-class (not all apparent on this plot).  Histograms above and below show the marginalized distributions that have been rescaled to represent posterior probability density functions. They are normalized to integrate to unity, then multiplied by the SN sub-class priors based on the host galaxy stellar population (row b in Supplementary Table 2).  Open markers show the observed photometry of the SN. Dotted vertical lines mark the magnification correction based on the preferred LENSTOOL model (model E, described in Methods: Lens Modeling).  Closed markers show the resulting magnification-corrected photometry, with asymmetric error bars reflecting the systematic uncertainty derived from the five lens model variants.  Horizontal error bars in the upper panel indicate the observed uncertainty in the SN color (not affected by lensing).  The relevant SN photometry markers are repeated in the histogram side-panels with arbitrary vertical positions.  All three SN images are located in regions of color-magnitude space that are expected to be dominated by Type Ia SN.}
    \label{fig:my_label}
\end{figure}

\begin{figure*}
    \centering
    \includegraphics[width=0.8\textwidth]{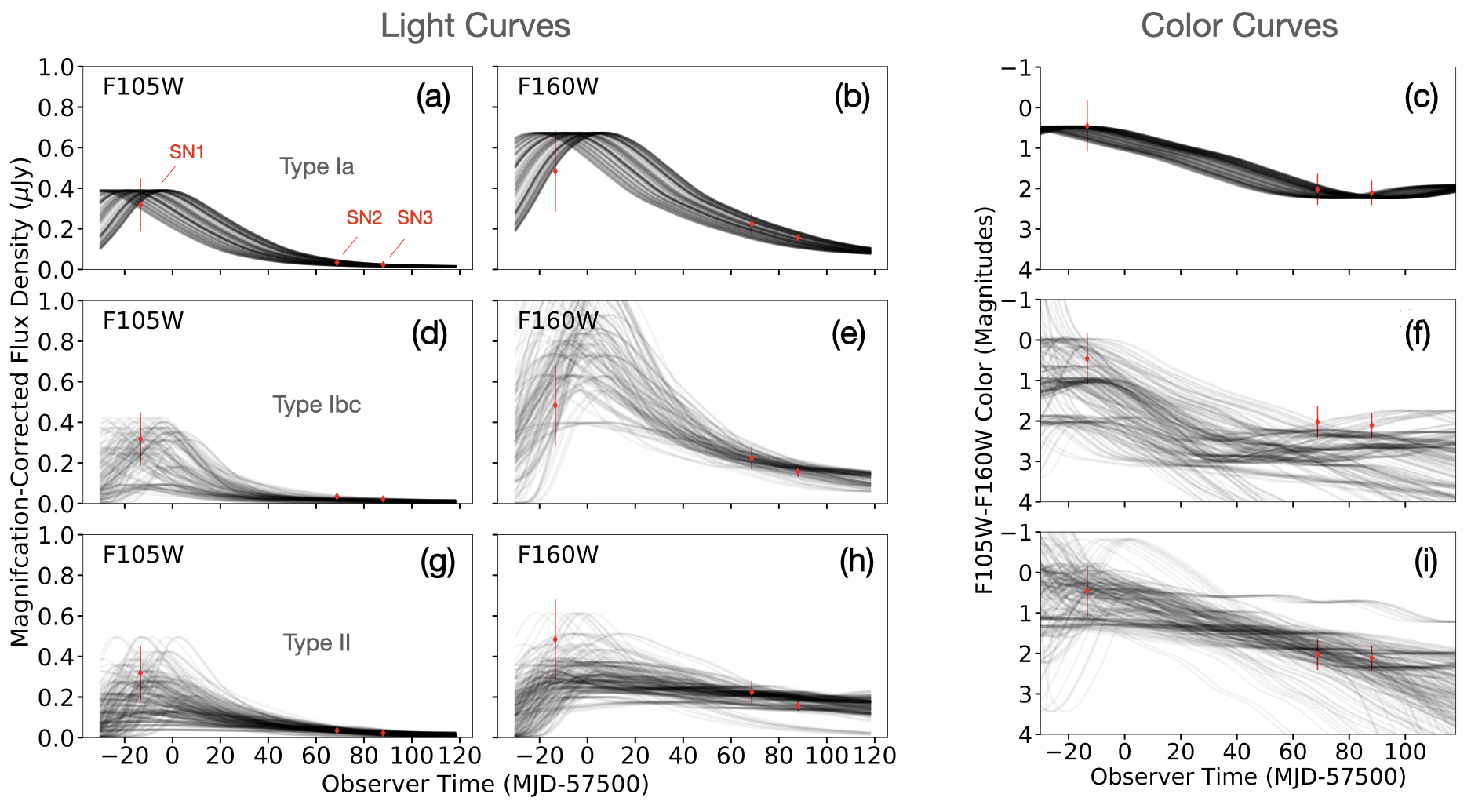}
    \caption{{\bf A representative set of light curve and color curve models from the STARDUST2 classification algorithm.} 
     Panels a-f show F105W and F160W, as indicated, plotting the model light curves in black and photometry as red markers.  Panels g-i show the F105W-F160W color curves and color data. All data points as shown have been corrected for magnification and shifted in time using the preferred LENSTOOL model (model E, described in Methods: Lens Modeling). Plotted error bars include the measurement uncertainty and the lens modeling magnification uncertainty. Data points in the right column also include this magnification uncertainty, even though cluster-scale lensing is achromatic, because the STARDUST2 analysis was done on the light curve data, not the color data directly. In all panels the first data point is SN image 2, followed by image 1 and image 3. In each panel the black curves show 200 SN light curve models drawn at random from the nested sampling sequence of the STARDUST2 (sncosmo) classification.  }
    \label{fig:my_label}
\end{figure*}

\begin{figure}
    \centering
    \includegraphics[width=\textwidth]{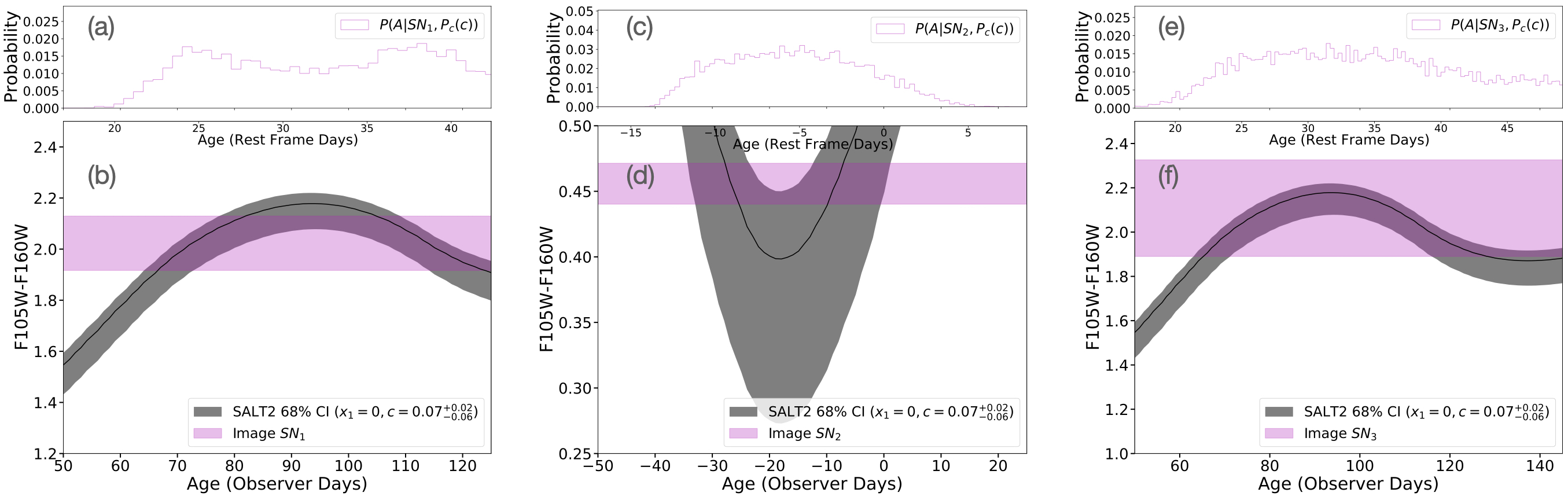}
    \caption{{\bf Color-based age constraints for AT 2016jka.}
    Constraints are shown separately for Image SN1 (panels a and b), image 2 (c and d), and image 3 (e and f), using the methodology described in Methods: Color Curve Age Constraints. Large lower panels (b, d and f) show the observations and model fits.  Each magenta shaded region shows the 1$\sigma$ range of the measured F105W-F160W color, which corresponds to a U-V color in the rest-frame.  The model fits are shown as grey shaded regions, indicating the 68\% confidence interval of the best-fit SALT2 color curve, with the median model shown as a solid line. The small upper panels (a, c and e) show the posterior for the age of each image from SNTD, using a prior on the SALT2 color parameter (c) based on known population characteristics of SNIa. The effect of adding this prior is slight, with no significant deviation from the best-fit value of $c=0.02^{+0.04}_{-0.05}$).
    }
    \label{fig:my_label}
\end{figure}

\begin{figure}
    \centering
    \includegraphics[width=0.8\textwidth]{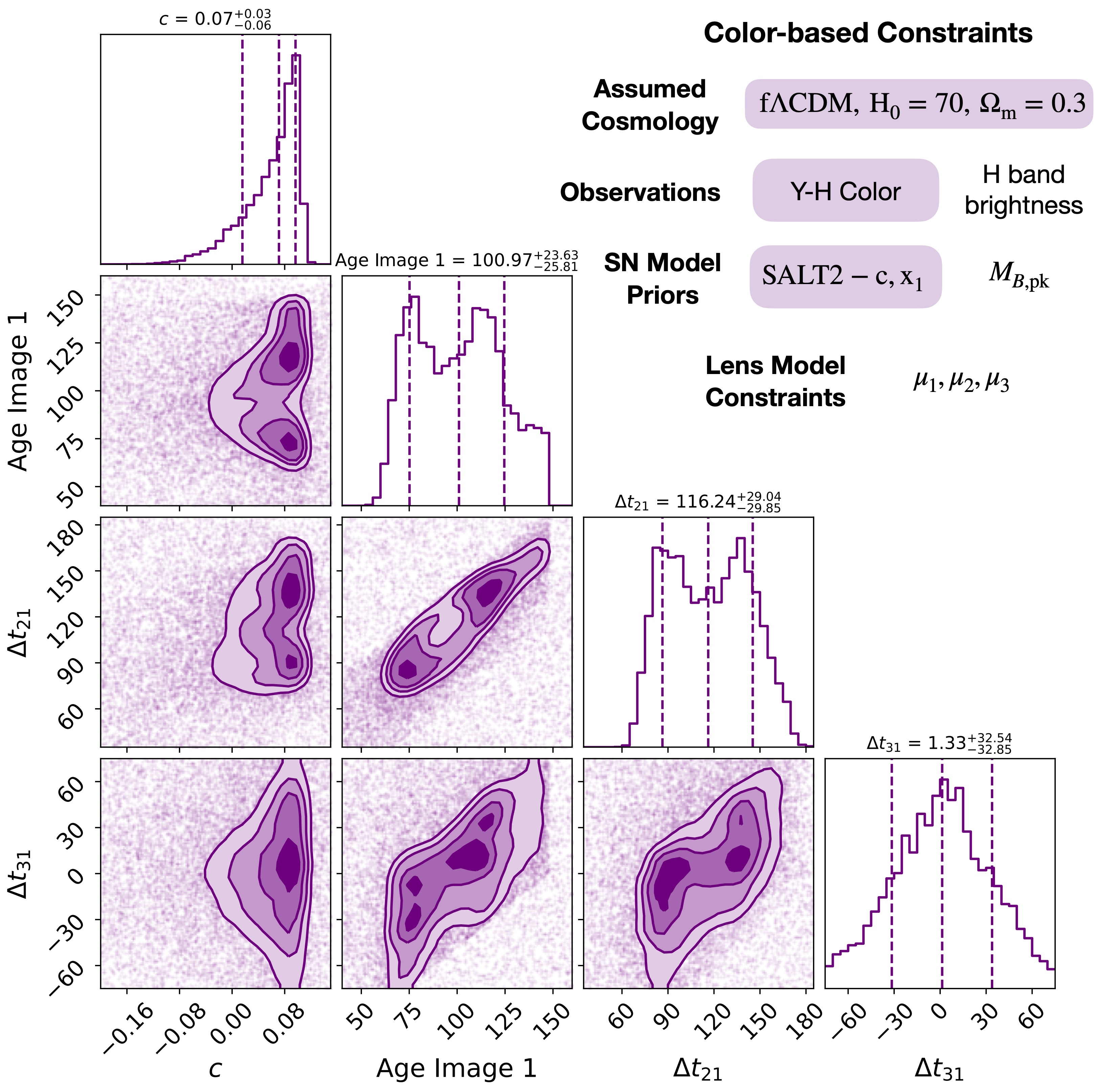}
    \caption{
    {\bf Marginalized and joint posterior distributions for the color curve age constraints measured in this analysis.}
    Two-dimensional plots show MCMC sampling points as discrete dots, with contours for the high density regions, drawn at 0.5, 1, 1.5 and 2$\sigma$ levels.  Marginalized (1D) distributions are shown at the top of each column, with dashed vertical lines at the mean and $\pm1\sigma$ (marking 16\%, 50\% and 84\% levels).  We use a weak prior on the SALT2 color parameter (c), and set the SALT2 stretch parameter ($x_1$) to 0. This method is fully independent of lens modeling.  The table in the upper right lists all priors, observations, and lens model information used for SN age estimates in this work.  Only the highlighted components were used for the constraints shown here.
    }
    \label{fig:my_label}
\end{figure}

\begin{figure}
    \centering
    \includegraphics[width=\textwidth]{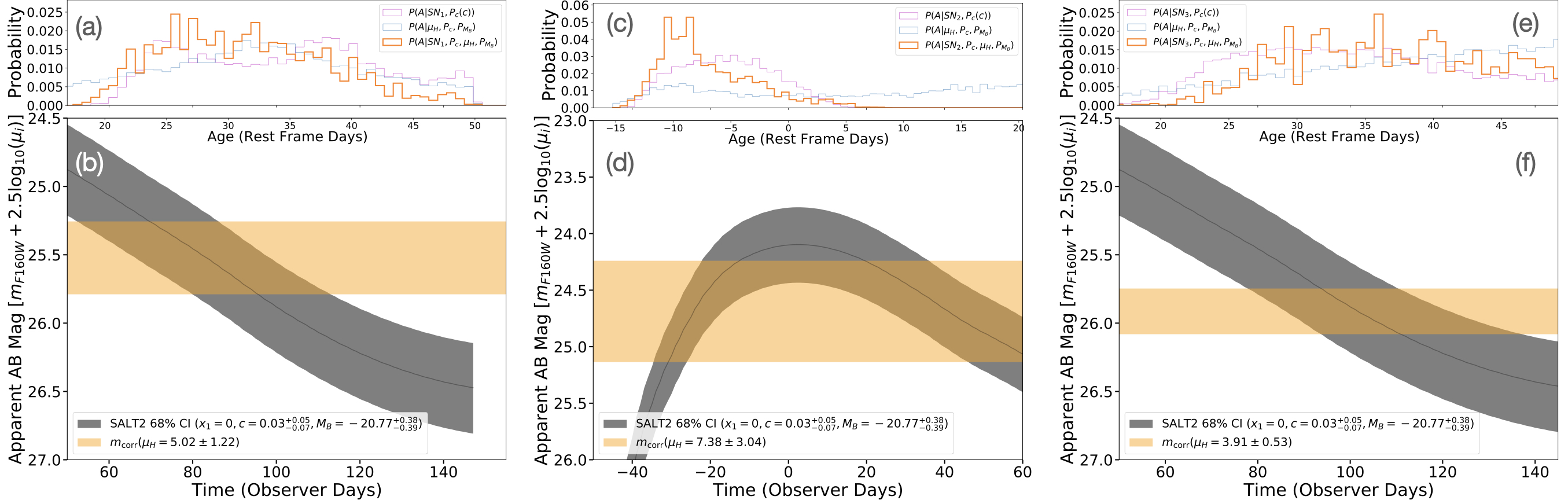}
    \caption{
{\bf Light-curve-based age constraints for AT 2016jka.}
Constraints are shown separately for Image SN1 (panels a and b), image 2 (c and d), and image 3 (e and f), using the methodology described in Methods: Light Curve Age Constraints. Large lower panels (b, d and f) show the observations and model fits. Each orange shaded region shows the 1$\sigma$ range of the measured F160W magnitude after lens model correction (see Table 1), which corresponds to roughly V band in the rest-frame.  The model fits are shown as grey shaded regions, indicating the 68\% confidence interval of the best-fit SALT2 light curve, with the median model shown as a solid line.  Small upper panels (a, c and e) show the posterior distributions from SNTD for the age of each image that is independent of the lens model (magenta, same as Extended Data Figure 7), using the preferred lens model E (light blue), and the combination of both methods (orange).
    }
    \label{fig:my_label}
\end{figure}

\begin{figure}
    \centering
    \includegraphics[width=0.8\textwidth]{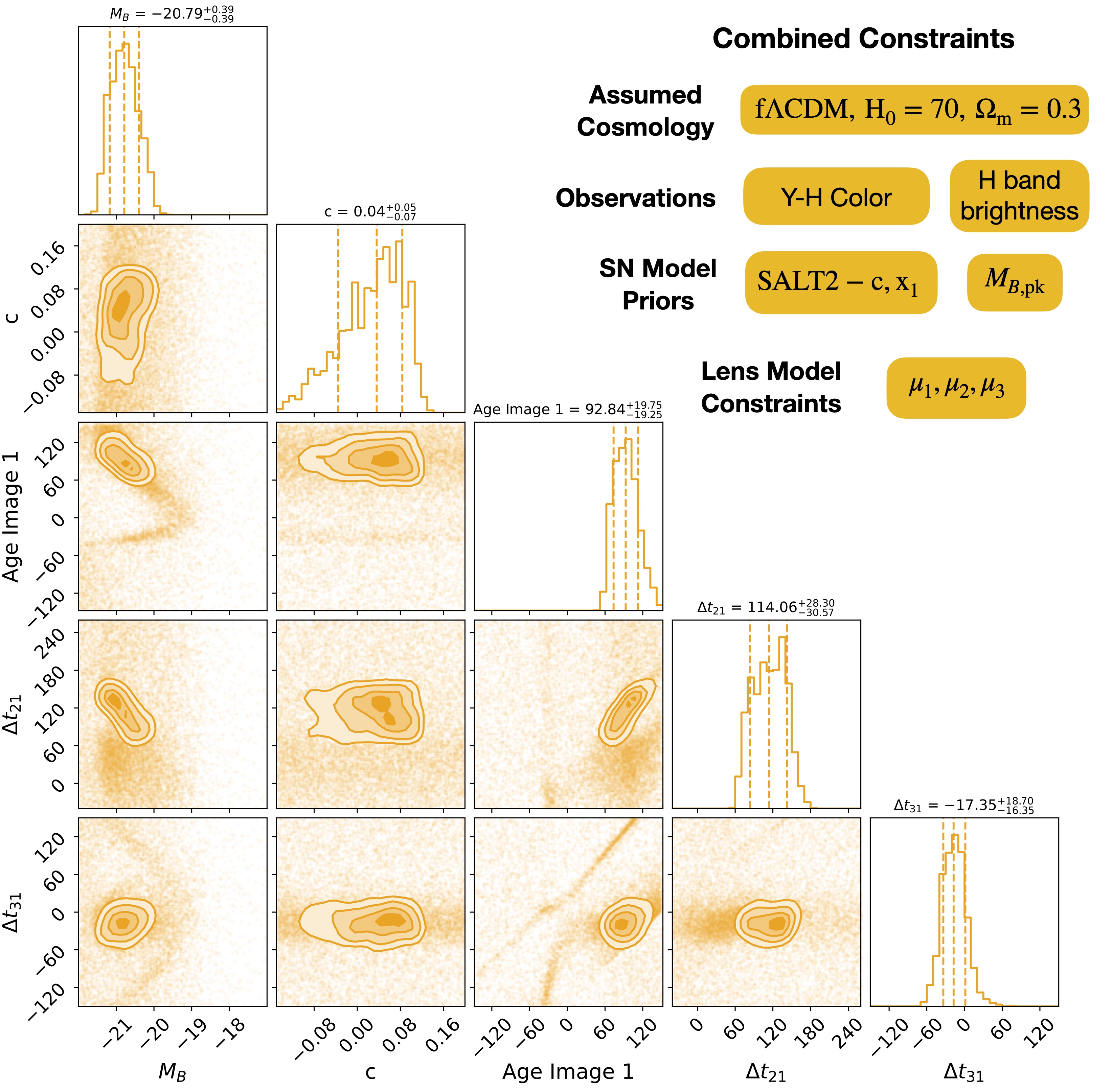}
    \caption{{\bf Marginalized and joint posterior distributions for the final age constraints measured in this analysis.} 
    Two-dimensional plots show MCMC sampling points as discrete dots, with contours for the high density regions, drawn at 0.5, 1, 1.5 and 2$\sigma$ levels.  Marginalized (1D) distributions are shown at the top of each column, with dashed vertical lines at the mean and $\pm1\sigma$ (marking 16\%, 50\% and 84\% levels).  We use the color curve posterior as the prior for light curve fitting with lens model E, and include weak priors on the absolute magnitude of a SNIa (MB) and the SALT2 color parameter ($c$), and set the SALT2 stretch parameter ($x_1$) to 0.
    }
    \label{fig:my_label}
\end{figure}
\clearpage

\begin{center}
    \bigskip
    {\Huge \bf Methods}
    \bigskip
\end{center}

\twocolumngrid

\subsection*{Observations} %

The observations used in this work are summarized in Supplementary Table 1.   We processed all HST observations using the {\tt Drizzlepac} software utilities \cite{gonzaga_drizzlepac_2012}, aligned to a common astrometric reference frame and resampled to 0.1 arcseconds per pixel.  We then identified isolated and unsaturated stars in each image and used them to create an effective point spread function (ePSF) with $4\times$ oversampling, using the {\tt photutils} package from the {\tt astropy} software suite \cite{the_astropy_collaboration_astropy_2018}.   
To measure the SN photometry we followed two tracks.  As our primary method we performed ePSF fitting directly on the F105W ($Y$ band) and F160W ($H$ band) images where the SN was apparent.  This fitting allowed for a constant background flux to account for both the sky brightness and the background light of the cluster and host galaxy.  
As a second approach, we created ``pseudo-difference images'' by re-scaling the F110W and F140W images collected in 2019 (in which the SN is not present).  The transmission functions of the F110W and F140W filters are broader than F105W and F160W, and do not strictly overlap in wavelength.  The optimal scaling factor to produce a clean subtraction therefore depends on the spectral energy distribution of the source.  We set the scaling to 0.62 and 1.17 for F110W-to-F105W and F140W-to-F160W, respectively.  These values produced visually clean subtractions of the SN host galaxy MRG0138---meaning that they minimize the residual flux in the pseudo-difference images.  We then performed ePSF fitting on the SN in each pseudo-difference image, as before.   Both sets of photometry agree to within one standard deviation.  The reported photometry in Supplementary Table~2 are the measurements from the first method (collected directly from the un-subtracted images).  Also reported in Supplementary Table~2 are flux densities and uncertainties measured within $D=0.^{\prime\prime}7$ circular apertures at the position of the SN3 image in the 2019 \textit{HST} visits where the SN is not detected.

\subsubsection*{VLT Spectroscopy}
\label{sec:vltmuse}

We make use of integral field spectroscopic data obtained on the cluster core of MRG0138 with VLT/MUSE, publicly 
available as part of the program 0103.A-0777(A) (PI: Edge). Three exposures of 970\,sec each were taken with a small dithering offset and 90 degree rotations in between. This dataset was reduced and analysed using the MUSE data reduction pipeline v.2.7 \cite[ref][]{weilbacher_data_2020} for basic calibration (bias, flat-field, wavelength, LSF, geometry) as well as flux calibration, sky subtraction and astrometry.  We  also make use of the self-calibration technique \cite{bacon_muse_2017} to remove illumination systematics, specifically tuned for the case of crowded fields in the central region of galaxy clusters (Richard et al. in prep.). The combined datacube is then processed through {\tt ZAP} \cite{soto_zap_2016} which applies a PCA technique to remove sky subtraction residuals. The final datacube covers the central 1x1 arcmin$^2$ around the cluster center with 0.2\arcsec$\times$0.2\arcsec$\times$1.25\AA\ pixels.

We have extracted spectra for each HST detected source and inspected them for redshift measurements. In addition, we have run the {\tt muselet} software (part of the MPDAF package \cite{piqueras_mpdaf_2019};  \href{https://mpdaf.readthedocs.io/en/latest/muselet.html}{mpdaf.readthedocs.io}) to search for line emitters not directly associated with HST sources \cite{mahler_strong_2018,lagattuta_probing_2019}. Apart from the lensed quiescent galaxy, we measured spectroscopic redshifts for cluster members and one ring-like background galaxy north of the BCG at $z=0.766$.  Finally, we have measured the velocity dispersion of the BCG to be 390$\pm$10 km s$^{-1}$.  This measurement was not available for use in our blind lens modelling, but could provide a useful constraint for future lens model improvements (Supplementary Note: Future Work).

\subsection*{Lens Modelling}

To model the mass distribution in the MACS J0138.0-2155 cluster core we use the latest version of \lenstool \cite{jullo_bayesian_2007} (\href{ https://git-cral.univ-lyon1.fr/lenstool/lenstool}{git-cral.univ-lyon1.fr/lenstool}), which performs a Bayesian analysis with an MCMC sampler to estimate the best fit and uncertainty on each parameter of the mass distribution. 

The strong-lensing constraints used are the locations of multiple images found in HST and MUSE/VLT. We group them into 3 systems: (a) the 4 images of the quiescent galaxy hosting MRG0138-SN, (b) the 3 observed images of \SNABC assumed to be at the same redshift, and (c) the diffuse arc-like structure identified in HST and confirmed as an [O\textsc{ii}]$\lambda$3727 emitter in the VLT/MUSE datacube (see previous section). The image coordinates and redshifts used in the lens model are summarised in Supplementary Table 3.

The cluster mass modelling is performed similarly to other massive strong lensing clusters observed with HST \cite{richard_mass_2014}. The total mass distribution is parametrized as a combination of multiple dPIE (double Pseudo Isothermal Elliptical) profiles describing both cluster-scale and galaxy-scale dark matter haloes. dPIE are elliptical isothermal profiles with both a core radius and a cut radius where the density flattens and drops, respectively. In the case of MRG0138 the mass distribution is dominated by a single mass concentration centered on its Brightest Cluster Galaxy (BCG). We therefore use a single cluster-scale halo at a fixed cut radius of 1\,Mpc.  We add a single galaxy-scale halo on each cluster member, where the shape parameters (halo center, ellipticity) are fixed to their measured HST morphology and their core radius is negligible (fixed at 0.15\,kpc). 
Extended Data Fig.~1 shows the locations of all components of the cluster model, including 32 cluster members. Cluster members were identified by the combination of red sequence selection (based on the F814W$-$F160W colour) and MUSE spectroscopy.

The majority of cluster members are elliptical galaxies selected from the red sequence, and to reduce the number of parameters we assume they follow the scaling relations: $\sigma=\sigma^*\ {\Big(\frac{L}{L^*}\Big)}^{(1/4)}$ for the velocity dispersion, 
and $r_{\rm cut}=r_{\rm cut}^*\ {\Big(\frac{L}{L^*}\Big)}^{(1/2)}$, assuming the Faber-Jackson relation and a constant M/L ratio respectively. $\sigma^*$ and $r^*_{\rm cut}$ are model parameters for a cluster member at the characteristic luminosity $L^*$.  Following the discussion in \cite{richard_locuss_2010} we fix $\sigma^*=158$ km/s and $r_{\rm cut}^*$=45 kpc. 

We individually optimise the $\sigma$ and $r_{\rm cut}$ parameters for 4 specific galaxies which are not expected to follow the aforementioned scaling relations: the BCG and three perturbers P1 to P3. These perturbers are either blue gas-stripped galaxies infalling into the cluster core, and/or located very close to the images of the \SNABC host galaxy, perturbing its apparent morphology with additional lensing. This choice of perturbers is similar to the ones used in the model by \cite{newman_resolving_2018}. 

\lenstool optimises the parameters of the model by minimising the overall root mean square dispersion (RMS) between the predicted and observed locations of the multiple images. The best fit parameters of each mass component are provided in Supplementary Table~4. Uncertainties reported there are derived from the MCMC models, by sampling their posterior probability distribution. 

We developed five lens model variants blindly (i.e., without knowing the impact of each lens model variation on the transient classification or time delay inferences). Model A was the first viable model developed, which did not include additional perturbers, and did not include the additional lensed background source at $z=0.7663$.   In Model B we allowed for the location of the main cluster dark matter halo to be free, with an offset from the reference position taken at the BCG center.  Model C allowed the same central position offset and also relaxed the constraints on the BCG ($\sigma$ and $r_{\rm cut}$).  Model D fixed the primary dark matter halo at the BCG center, but still relaxed the constraints on the BCG $\sigma$ and $r_{\rm cut}$.
The final model, and the one selected as the preferred model prior to unblinding, is model E, which includes all four perturbers described above, and includes the additional background object at $z=0.7663$.
Note that all of the lens model variants A-D would give a slightly larger magnification for all  three SN images.  This means that our estimated systematic uncertainties are one-sided, as can be seen on Fig.~\ref{fig:class} and Extended Data Fig.~2.

This model is then used to predict the magnification and time delays for the three observed images of \SNABC, as well as the location of the fourth and fifth images, which are still to appear. 
These predictions are summarised in Table \ref{tab:time_delays}, but note that the magnification predictions for image SN4, as well as all predictions for SN5, should be treated with caution, as the lens models presented here have known shortcomings.  For example, the best-fit velocity dispersion for the BCG is 700 km/s (Supplementary Table~4), though we have measured this property from MUSE spectra to be 390$\pm$10 km/s. Further lens modelling is needed to incorporate such additional constraints and to fully quantify potential systematic biases (Supplementary Note: Future Lens Modelling).

The final \lenstool model E reproduces all multiple system positions with an RMS of 0.15\arcsec.
The 1$\sigma$ uncertainty from model E for the SN4 location is the ellipse overlayed on panels {\it e} and {\it i} in Fig.~\ref{fig:layout}. As the lens model reproduces the location of the SN images within a small uncertainty, these predictions are computed with \lenstool using the barycenter of all source positions corresponding to images SN1, SN2 and SN3 as the same reference source position.
Comparing our lens modelling to the previous  model of this cluster from \cite{newman_resolving_2018}, we find that the magnification estimates are broadly consistent, though systematically lower (Supplementary Note: Comparison to Previous Lens Modelling).

\subsection*{Classification}

The lens modelled time delays between the images are $\sim100$ observer-frame days, 
but we see that three images of the transient are visible simultaneously.
From this we can infer that the visibility time of the transient in the $z=1.95$ rest-frame must be at least $\sim$30 days. 
Similarly, with expected magnifications in the vicinity of $\mu\sim10$, the measured apparent magnitudes near 23 AB mag translate to a rest-frame absolute magnitude near $M_B \sim-19.5$ mag (too bright to be a nova, luminous blue variable, or other low-luminosity stellar transient).
Taken together, these indicators strongly suggest that the transient is a supernova (SN). 

Although we have invoked the lens model in this analysis, we note that the inferences are not strongly dependent on the specific lens model predictions.  To make the observed transient images consistent with a fast or low-luminosity transient, the time delays and/or magnifications would have to be changed by more than a factor of 2.  In the analysis to follow, we will work under the assumption that \SNABC is a SN.

\subsubsection*{SN Sub-Classification Based on Host Galaxy}
With this transient identified as a SN, we now seek to identify the most likely SN type, under the assumption that it belongs to one of the three most common sub-classes (Ia, II, Ib/c).   We first use two methods that rely only on measured properties of the host galaxy to {\it circumstantially} infer the type. This inference is less strongly dependent on the lens model, helping to reduce any bias associated with a lens model-dependent classification. 

Although Type Ia SNe are found in all types of galaxies, CCSNe are limited to galaxies with relatively young stellar populations.  We can therefore infer some information about the SN type using the observed host properties combined with knowledge of the relative rates of Type Ia and CCSNe in different stellar populations \cite{mannucci_supernova_2005}.  In the case of the host galaxy MRG0138, we have a very well-constrained spectral energy distribution (SED) extending out to far-infrared wavelengths with \textit{Spitzer} IRAC data \cite{newman_resolving_2018,newman_resolving_2018-1}.  
From the SED fitting we derived the host galaxy's 
rest-frame $B-K$ colour and absolute magnitude, $M_K$, which serve as proxies for the stellar population age and have been empirically calibrated with SN rates in the local universe \cite{foley_classifying_2013}.  We  adopt a lensing magnification correction using \lenstool model E to get $M_K$. 
The $B-K$ colour is not affected by the foreground lens.
Using the {\tt galsnid} method \cite{foley_classifying_2013} we 
derive a $75\%$ probability that \SNABC is of Type Ia (Supplementary Table~5).
We used host galaxy image 2 for this purpose, with the flux-weighted harmonic mean magnification $\mu=8.3$ to derive $M_K$ (see Supplementary Table~6).
We also evaluated the other host galaxy images and found no change in the resulting SN classification probability.

As an alternative host galaxy classification constraint, we use the derived properties of the host galaxy stellar population directly,  rather than adopting colour and magnitude proxies. The MRG-M0138 galaxy has high mass ($\log_{10}(M/M_{\odot})=11.7$) but is a very quiescent galaxy, with a specific star formation rate of $\sim10^{-11.3}$ yr$^{-1}$  and a stellar population that is well-matched by an exponential star formation history with an age of $1.4$ Gyr \cite[ref][]{newman_resolving_2018}. 
The massive stars that end as CCSN explosions have main-sequence lifetimes of $<40$ Myr \cite[ref][]{smartt_progenitors_2009},  making it unlikely that CCSN progenitors make up a significant fraction of the MRG-M0138 stellar population---though the high total stellar mass makes it possible that pockets of young stars are present.  We define classification probabilities based on the projected SN rate for each SN sub-class, derived from the host galaxy's stellar mass and star formation rate \cite{li_rates_2012}.  This yields a 62\% probability that \SNABC is of Type Ia (Supplementary Table~5).

\subsubsection*{SN Sub-Classification Based on SN Photometry}

To improve the classification of the \SNABC sub-type, we now bring in observed photometry of the SN itself, and again we adopt two methods.  The first method uses only magnification information from the lens model, and the second uses both the modelled magnification and time delay predictions.  In both cases we adopt the stellar-population-based host galaxy classification probabilities as priors.

Extended Data Fig.~2 illustrates the first approach.  After applying the magnification corrections, each of the three images of the SN are mapped to colour-magnitude space.  We then treat each observed point (each SN image) separately, comparing their colour-magnitude location to a simulated population of unlensed SNe.  Our simulation uses the {\tt sncosmo} package \cite{barbary_sncosmo_2016}  to generate 10,000 SN for each of the three principal SN sub-classes (Ia, Ib/c, II), all at z=1.95.  
We then compute the number of simulated SN within a rectangular region around each observed point. The width of this sampling region is set to 3 times the observed colour uncertainty, and the height is equal to the lens-modelling magnification uncertainty. The $\mu$ uncertainty used here includes an estimate of the systematic uncertainty for each SN image, derived from the spread of magnifications across lens model variants (similar to the methodology of \cite{newman_resolving_2018}).  
Note, however, that the lens model variations evaluated here  do not vary the assumption of the density profile, which can strongly affect magnifications (see also the Supplementary Note: Comparison to Previous Lens Modelling).

We take the number of simulated SN for each type within the sampling region as an estimate of the likelihood that \SNABC belongs to that class.
Note that in this case there is no need to apply a cut to the simulated sample to account for detectability, because the $5\sigma$ limiting magnitude of our HST observations is 26.5 AB mag, and after accounting for magnification of $\sim$1.5 mag (Table 1) this becomes $m_{lim}\sim28$ mag. This means that all the points shown in Extended Data Fig.~2 (and therefore all simulated SN entering our classification counts) would be easily detectable in our HST imaging.  Multiplying by the prior probabilities derived from host galaxy properties, we finally derive the probability that the SN is of Type Ia as $p(Ia)=0.92$, 0.98, and 0.95 from the three SN images SN1, SN2 and SN3, respectively. Supplementary Table~5 reports the mean of our three classification probabilities for each sub-class. 

As a second photometric classification of this SN, we used the {\tt STARDUST2} Bayesian light curve classification tool \cite{rodney_type_2014}, which is also built on the underlying {\tt sncosmo} framework. Here we adopt both the predicted magnifications and time delays from the best lens model, which allows us to put the photometry from the three images together as a composite ``light curve'' and compare against simulated light curves.  {\tt STARDUST2} uses the {\it SALT2-extended} model to represent Type Ia SN \cite{guy_salt2:_2007, pierel_extending_2018} and a collection of 42  spectrophotometric time series templates to represent CCSN (27 Type II and 15 Type Ib/c).  These CCSN templates comprise all of the templates developed for the Supernova Analysis software {\tt SNANA} \cite{kessler_snana:_2009}, derived from the SN samples of the Sloan Digital Sky Survey \cite{frieman_sloan_2008,sako_sloan_2008, dandrea_type_2010}, Supernova Legacy Survey \cite{astier_supernova_2006}, and Carnegie Supernova Project \cite{hamuy_carnegie_2006, stritzinger_he-rich_2009, morrell_carnegie_2012}.  With {\tt STARDUST2} we use a nested sampling algorithm to measure likelihoods over the SN simulation parameter space.  
Extended Data Fig.~3 shows the magnification- and time-delay-corrected photometry of \SNABC and a random sampling of light curve models from the {\tt sncosmo} nested sampling algorithm employed by {\tt STARDUST2}.
Nested sampling is a Monte Carlo method that traverses the likelihood space in a manner that samples the Bayesian likelihood \cite{skilling_nested_2004}.  These sample light curves and colour curves therefore give a visual representation of how well each SN sub-class (Ia, II, Ib/c) can match the observed data.
This figure shows that the limited photometric data can be reasonably well fit by at least one model from any of these three sub-classes. 
The density of curves in the top panels demonstrates that the Type Ia SN model is consistently a good match to the data. However, the range of model parameters that allow such a fit to the data is much more limited for the heterogeneous CCSN types. 
To compute the posterior probability distribution we adopt priors for each of the three SN classes, again using the classification probabilities derived from the \SNABC host galaxy stellar population properties (Supplementary Table~5). 
Marginalizing the posterior  probability distributions over all free parameters, we find a 94\% probability that \SNABC is of Type Ia (Supplementary Table~5).  

The combination of evidence from the derived host galaxy properties  and SN photometry supports the conclusion that \SNABC  is  a Type Ia SN with $>90\%$ confidence.  Although we have adopted some lens modelling corrections for all of these methods, this conclusion is not sensitive to the choices we can reasonably make for modelling the lens.  As shown in Extended Data Figure~4, every  \lenstool model we have evaluated locates the \SNABC  de-magnified position within the region of colour-magnitude space dominated by Type Ia SN. Dust is also not a confounding factor here. 
The simulations used in both SN-based classification methods include dust extinction at the source plane.  If a significant screen of dust exists in the lens plane, this would have the effect of making the SN appear dimmer and more red, so correcting for that extinction would move the \SNABC points upward and leftward on Extended Data Fig.~2, which would not shift it into the regions occupied by Type II and Ib/c SN.

\subsection*{Time Delay Estimation}

We constrain the relative time delays of \SNABC by using two separate methods to estimate the age of the SN at each image during the single observed epoch. The preferred method of SN time delay measurements involves measuring the time of peak brightness for the SN at each image by fitting the light curves, and taking the difference between each measurement as the relative time delay \cite{pierel_turning_2019,dhawan_magnification_2019,huber_strongly_2019}. With only a single observed epoch, this method is impossible due to model parameter degeneracies, and we must rely on colour and brightness to constrain the age of each image of the SN. Such age estimates are sometimes referenced to the time of explosion, but in this case we use the observer's convention, setting age=0 as the time of peak brightness in the rest-frame B band ($\lambda\sim4500$ \AA).  Each of these images stems from the same SN explosion, so the difference between the measured age of each image is also a measure of the relative time delay.

In all of the light curve fitting exercises described below, we also fix the SALT2 ``stretch'' parameter at $x_1=0$. This parameter defines the shape of the SN Ia light curve (the rate of decline in brightness). If we  allow $x_1$ to be a free parameter, there is no useful constraint on it.  It is highly degenerate with the time delay between the images, which is of course a free parameter in all the fits.  
As a check, we have also tried fixing $x_1$ to other values from $-1$ to $+1$, and the time delay results change by less than 5 days, which is well within all error bars.   
Fixing $x_1$ in this way is comparable to the analysis that will be possible in the 2030s when the fourth SN image is observed. We expect that a SALT2 fit to a well-sampled light curve from the fourth image will provide a tight constraint on $x_1$.  That measured $x_1$ will then be propagated back as a fixed parameter (with small uncertainty) into revised fitting of the SN images 1-3. Because of this, we do not incorporate the $\pm$5 day systematic uncertainty in the time-delay errors reported in Table~\ref{tab:time_delays}.

\subsubsection*{Colour Curve Age Constraints}

We first attempt to constrain the relative time delays using the colour of each observed image, which is independent of the lens model and possible because the phenomenon of gravitational lensing is intrinsically achromatic. One important caveat to this principle is that {\it microlensing} effects are not generally achromatic, because the microlensing caustics may cause differential magnification on the scale of the SN radius \cite{goldstein_precise_2018,foxley-marrable_impact_2018,bonvin_impact_2019}.  Hence, if the expanding SN shell has a colour gradient then microlensing may introduce spurious features in the observed colours of the SN \cite{kochanek_quantitative_2004,vernardos_joint_2018}.   Ref.~\cite{goldstein_precise_2018} found that such chromatic microlensing is most likely not present for lensed Type Ia SNe in the period up to about 25 rest-frame days after explosion ($\sim$15 observer days after peak brightness for \SNABC). Only image 2 is likely in the achromatic microlensing phase, but Ref.~\cite{goldstein_precise_2018} found extremely small deviations in the rest-frame $U-V$ colour curve due to microlensing at the 68\% confidence interval, and up to a $\sim0.2-0.4$ mag difference with 99\% confidence. While such extreme microlensing could alter the results for images 1 and 3, it would not alter the measurement of image 2 as it is likely in the achromatic phase.
Fortuitously, images 1 and 3 are minima, which are less susceptible to high deviations owing to microlensing when compared to image 2, which is a saddle.

We use version 2 of the SuperNova Time Delays (SNTD) package (publicly available at \href{https://github.com/jpierel14/sntd/releases/tag/2.0}{github.com/jpierel14/sntd}) with documentation at 
\href{https://sntd.readthedocs.io/en/latest/}{sntd.readthedocs.io}, which has several improvements over the original SNTD package \cite{pierel_turning_2019}. The SNTD package employs a nested sampling algorithm within three separate methods to measure time delays, and is designed to fully utilize the information present in SN light curve templates \cite{hsiao_k_2007,guy_salt2:_2007,kessler_results_2010,pierel_extending_2018} to reduce the impacts of microlensing and make more accurate measurements. We use the ``colour'' method present in SNTD, which attempts to reconstruct the intrinsic colour curve using the SALT2 model as a template \cite{guy_salt2:_2007}. This method fits the age of each image simultaneously, while also varying the SN model parameters.  This means we are finding a single set of SN model parameters to describe the intrinsic photometric evolution of the SN, and also finding the age (time from peak brightness) for each of the three images. 

The colour curve constraints resultiing from this process are shown in Extended Data Fig.~4.  Joint and marginalized posterior distributions from SNTD for the SALT2 SN model parameters and the measured ages for each image are shown in Extended Data Fig.~5.  
In Extended Data Fig.~4, one can see the measured colours intersect the model at two distinct locations for images 2 and 3 of \SNABC---meaning there are two plausible ages.  
This results in a double peaked posterior distribution in Extended Data Fig.~5. 
This is caused by a model parameter degeneracy that could be broken in a way independent of the lens model if a sufficiently precise colour curve of image 4 is obtained in the future.

\subsubsection*{Light Curve + Lens Model Age Constraints}

In order to break the age degeneracies in the colour-based constraints using only data available today, we need to use some information about the relative brightnesses of the \SNABC images.  
For this step we can no longer be independent of the lens models, as we must use the lens-model-predicted magnification values to de-magnify the observed photometry for comparison to SN models (note, however, that we again do not use any time delay information from the lens model for this method).

For the five lens models (A-E) described above, we correct the observed flux density of each image (in both F105W and F160W bands) using the predicted lensing magnification ($\mu$). 
Next we employ SNTD's ``series'' method, as it is most effective for sparse sampling, to attempt a reconstruction of the intrinsic SN light curve \cite{pierel_turning_2019}. 
Once again the age of each image is constrained simultaneously, while also varying the SN model parameters. At this stage we adopt weak priors on the intrinsic Type Ia SN luminosity \cite{wang_determination_2006} and Type Ia SN colour \cite{mosher_cosmological_2014} to help break degeneracies in the light curve model.  Extended Data Fig.~6 shows the resulting 
light-curve-based constraints on the age of each SN image. 

As our final method to incorporate the colour and brightness information together, we use the colour-based posterior probability distributions (Extended Data Fig.~4) as priors for 
the light-curve based constraints. 
In this approach, we must use only a single photometric band for the light curve constraint, so that we are not ``double-counting'' the colour information by simultaneously fitting to two bands together.  
We adopt F160W as the single band, since it is close to the rest-frame V band, where the SALT2 Type Ia SN model is very well constrained.  The joint posterior distributions from this method are shown in Extended Data Fig.~7.

\clearpage
\onecolumngrid

{\bf Data Availability Statement:} All HST images used in this work are available from the Mikulski Archive for Space Telescopes (mast.stsci.edu). HST data from 2016 when the transient was active are from the program HST-GO-14496 (\url{archive.stsci.edu/proposal_search.php?id=14496&mission=hst}).
Data collected in 2019 are from the REQUIEM program, HST-GO-15663 (\url{archive.stsci.edu/proposal_search.php?id=15663&mission=hst}).  All VLT MUSE spectroscopic data used in this work are available from  the ESO Archive Science Portal (archive.eso.org). The MUSE datasets can be found at 
\url{archive.eso.org/dataset/ADP.2019-10-07T18:14:24.762}, \url{archive.eso.org/dataset/ADP.2019-10-07T18:14:24.751}, and \url{archive.eso.org/dataset/ADP.2019-10-07T18:14:24.776}.
All derived data supporting the findings of this study (photometry, lens model inputs, etc.) are available within the paper and its supplementary information files.  
\medskip

{\bf Code Availability Statement:} 
All software tools used in the analysis are publicly available, as indicated in the text.  The software used for figure creation, including input data files, can be downloaded from \url{github.com/gbrammer/mrg0138_supernova}.
\medskip

Correspondence and requests for materials may be addressed to S.R. (\href{mailto:srodney@sc.edu}{srodney@sc.edu}) and G.B. (\href{mailto:gabriel.brammer@nbi.ku.dk}{gabriel.brammer@nbi.ku.dk}).

\clearpage

\renewcommand{\tablename}{Supplementary Table} 
\renewcommand{\figurename}{Supplementary Figure} 
\setcounter{table}{0}
\setcounter{figure}{0}

\begin{center}
\bigskip
\vspace{0.5in}
{\Huge \bf Supplementary Material}
\end{center}
\vspace{1in}

\begin{table}[htb]
\centering
\begin{tabular}{ccccccr}
    \multicolumn{1}{c}{\bf Telescope} & \multicolumn{1}{c}{\bf Instrument} & \multicolumn{1}{c}{\bf Band} & \multicolumn{1}{c}{\bf UT Date} & \multicolumn{1}{c}{\bf MJD} & \multicolumn{1}{c}{\bf Exp. Time [s]}\\
\midrule
\textit{Spitzer} & IRAC      & $3.6\,\mu\mathrm{m}$      & 2016-03-15 03:44:04 & 57462.156 &   212 \\ 
\textit{Spitzer} & IRAC      & $4.5\,\mu\mathrm{m}$      & 2016-03-15 03:44:04 & 57462.156  &  241 \\
\textit{HST}     & ACS/WFC   & F555W                     & 2016-06-03 21:50:43 & 57542.910  &  5214 \\
\textit{HST}     & WFC3/IR   & F160W                    & 2016-07-18 23:14:50 & 57587.969$^*$ & 1611 \\
\textit{HST}     & WFC3/IR   & F105W                     & 2016-07-19 00:43:47 & 57588.030$^*$ & 3611 \\ 
\textit{Spitzer} & IRAC      & $3.6\,\mu\mathrm{m}$      & 2016-10-13 14:35:13 & 57674.608 &  468 \\ 
\textit{Spitzer} & IRAC      & $4.5\,\mu\mathrm{m}$      & 2016-10-13 14:35:13 & 57674.608 &  581 \\
\midrule
\textit{HST}     & WFC3/IR   & F110W                     & 2019-07-13 20:53:16 & 58677.870 &  706 \\ 
\textit{HST}     & WFC3/IR   & F140W                     & 2019-07-14 22:16:01 & 58678.928 &  353 \\ 
\textit{HST}     & WFC3/IR   & F125W                     & 2019-07-19 21:27:30 & 58683.894 &  706 \\ 
\textit{HST}     & WFC3/UVIS & F814W                     & 2019-07-21 18:50:53 & 58685.785 &  912 \\ 
\textit{HST}     & WFC3/UVIS & F390W                     & 2019-07-21 19:01:06 & 58685.792 &  1272 \\ 
\textit{HST}     & WFC3/IR   & F140W                     & 2019-07-21 22:42:22 & 58685.946 &  353 \\ 
\textit{VLT}     & MUSE      & 0.4--0.9$\,\mu\mathrm{m}$ & 2019-09-06 03:56:25 & 58732.164 &  2649  \\
\end{tabular}
\caption{\textbf{Record of MACSJ0138 observations used in this work.}  
Observations in which three images of the SN were detected are marked with an '*' in column five.
\label{tab:observations}
}
\end{table}

\begin{table}[htb]
\centering
\begin{tabular}{cccc}
Image & Obs. Date (MJD) & Filter & Flux density [$\mu$Jy] \\
\midrule
SN1 & 57588.03 & F105W ($Y$) & 0.18  $\pm$ 0.02 \\
SN2 & 57588.03 & F105W ($Y$) & 2.35  $\pm$ 0.02 \\
SN3 & 57588.03 & F105W ($Y$) & 0.09  $\pm$ 0.02 \\
SN1 & 57587.97 & F160W ($H$) & 1.13  $\pm$ 0.04 \\
SN2 & 57587.97 & F160W ($H$) & 3.57  $\pm$ 0.05 \\
SN3 & 57587.97 & F160W ($H$) & 0.61  $\pm$ 0.04 \\
\midrule
SN3 & 58677.87 & F110W ($Y$) & 0.01  $\pm$ 0.02 \\
SN3 & 58678.93 & F140W ($JH$) & 0.02  $\pm$ 0.02 \\
SN3 & 58683.89 & F125W ($J$) & 0.02  $\pm$ 0.03 \\
\end{tabular}
\caption{\textbf{Photometry of \SNABC.}  The final three rows indicate ``empty'' aperture flux densities measured at the position of image SN3 in the 2019 \textit{HST} visits.
\label{tab:photometry}}
\end{table}

\begin{table}[ht]
    \centering
    \begin{tabular}{cccl}
     ID &   R.A. (deg) & Dec. (deg) & $z$ \\
     \midrule
H1 & 24.5099018 & $-$21.9260130 & 1.95 \\
H2 & 24.5132090 & $-$21.9299032 & 1.95 \\
H3 & 24.5164138 & $-$21.9303172 & 1.95 \\
H4 & 24.5176117 & $-$21.9233433 & 1.95 \\
     \midrule
SN1 & 24.5151253 & $-$21.9306659 & 1.95 \\
SN2 & 24.5123198 & $-$21.9297875 & 1.95 \\
SN3 & 24.5100753 & $-$21.9273418 & 1.95 \\
     \midrule
3.1 & 24.5169659 & $-$21.9234814 & 0.7663 \\
3.2 & 24.5151039 & $-$21.9231406 & 0.7663 \\
3.3 & 24.5184361 & $-$21.9264250 & 0.7663 \\
3.4 & 24.5146833 & $-$21.9261020 & 0.7663 \\
    \end{tabular}
    \caption{\textbf{Multiple images used as constraints in our parametric lens model.} From left to right: image ID, right ascension, declination, spectroscopic redshift.  Coordinates are in the J2000 reference frame.}
    \label{tab:mulimages}
\end{table}

\begin{table}[htb]
    \centering
    \begin{tabular}{c|c|c|c|c|c|c|c|}
    Potential & $\Delta$R.A. & $\Delta$Dec. & $e$ & $\theta$ & $r_{\rm core}$ & $r_{\rm cut}$ & $\sigma$ \\
                  & [arcsec] & [arcsec] & & [deg] & [kpc] & [kpc] & [km\ $s^{-1}$] \\
\midrule
Cluster-DM & $ -0.7^{+0.4}_{-0.4}$ & $ -1.2^{+0.4}_{-0.4}$ & $ 0.81^{+0.02}_{-0.13}$ & $114.9^{+2.0}_{-4.1}$ & $31^{+13}_{-12}$ & $[1000]$ & $446^{+52}_{-70}$ \\
BCG            & $[  0.1]$ & $[ -0.1]$ & $[0.52]$ & $[-41.1]$ & $[0.15]$ & $136^{+42}_{-32}$  & $700^{+52}_{-57}$ \\
P1             & $[ 19.2]$ & $[-13.5]$ & $[0.49]$ & $[ 86.2]$ & $[0.15]$ & $[25]$            &  $152^{+30}_{-57}$ \\
P2             & $[ -5.0]$ & $[  6.9]$ & $[0.06]$ & $[  4.4]$ & $[0.15]$ & $[12]$            &  $23^{+111}_{-29}$ \\
P3             & $[ -0.8]$ & $[-16.7]$ & $[0.24]$ & $[-63.1]$ & $[0.15]$ & $[6]$             &  $110^{+35}_{-32}$ \\
L$^{*}$ galaxy &           &           &          &           & $[0.15]$ & $[45]$            & $[158]$\\
    \end{tabular}
    \caption{\textbf{Best fit model parameters for the MACS J0138 mass distribution.} From left to right: mass component, position relative to cluster center ($\Delta$R.A. and $\Delta$Dec.), dPIE shape (ellipticity and orientation), velocity dispersion, core and cut radius. The final row  is the generic galaxy mass at the characteristic luminosity L$^*$, which is scaled to match each of cluster member galaxies.  Parameters in square brackets are fixed {\it a priori} in the final model (version E). }
    \label{tab:massmodel}
\end{table}

\begin{table}[tb]
    \centering
    \begin{tabular}{lp{1.5in}cc|ccc}
        \multicolumn{1}{c}{Method} & \multicolumn{1}{c}{Data} & Lens info & Priors & p(Ia) & p(II) & p(Ib/c) \\
        \midrule
        a. Host color-mag & Host galaxy rest-frame $M_K$, $B-K$ & $\mu_{\rm host}$ & - & 0.75 & 0.19 & 0.06\\
        b. Host stellar pop. & Host galaxy mass \& star formation rate & $\mu_{\rm host}$ & - & 0.62 & 0.27 & 0.09 \\
        c. SN color-mag & SN F105W-F160W color, $m_{\rm F160W}$ & $\mu_{\rm SN}$ & b & 0.95 & 0.01 & 0.04\\
        d. SN light curve & F105W and F160W SN light curves & $\mu_{\rm SN}$, $\Delta t_{\rm SN}$ & b & 0.94 & 0.06 & $<$0.01 \\
    \end{tabular}
    \caption{\textbf{SN classification probabilities for \SNABC.} ``Lens info'' indicates the lensing information used to interpret or derive the 
    observational data: $\mu_{\rm host}$ and $\mu_{\rm SN}$ are the magnifications of the host galaxy MRG0138 and the SN, respectively; $\Delta t_{\rm SN}$ refers to the time delays between SN images 1, 2 and 3. In all cases the preferred \lenstool model E is used.  ``Priors'' indicates the host galaxy classification probabilities that were adopted as priors for the subsequent classification using SN data.}
    \label{tab:classification}
\end{table}

\clearpage

\subsection*{Supplementary Note: Comparison to Previous Lens Modeling}

\begin{table}[b!]
    \centering
    \begin{tabular}{crrrrr}
    Image     & $\langle$ $\mu_{\rm SN}$ $\rangle$ & $\mu_{\rm opt,SN}$ & $\mu_{\rm opt,gal.pk.}$ & $\mu_{\rm gal.avg.}$ & $\mu_{\rm N18,gal.avg.}$\\
\hline 
        1 & 3.9$\pm$0.5 &  6.7 & 4.35 & 10.0 & 12.5$\pm$5.4 \\
        2 & 7.4$\pm$3   & 15.2 &  6.9 &  8.3 & 10.3$\pm$3.1 \\
        3 & 5$\pm$1     &  4.3 & 3.64 &  4.2 & 4.9$\pm$1.6 \\
    \end{tabular}
    \caption{
    \textbf{Comparison of magnification predictions with the N18 lens model.}
    }
    \label{tab:mu_comparison}
\end{table}

Ref \cite{newman_resolving_2018} (hereafter N18) provides a detailed analysis of this cluster, including sophisticated modeling of the lens. 
The primary modeling in N18 uses a custom lens model code \cite{newman_luminous_2015} that traces the source Sersic profile(s) through to the image plane and fits the HST images directly. In addition, N18 also made a set of \lenstool models to estimate the uncertainties. Those models included both source and image plane optimization, and considered both generalized Navarro-Frenk-White (gNFW) \cite{zhao_analytical_1996} and dual pseudo isothermal elliptical (dPIE) density profiles for the cluster.   For this work, we used only \lenstool, only dPIE profiles, and considered only image-plane optimisation because it deals best with actual positional uncertainties.  We did not incorporate pixel-level flux data, because the \lenstool code can only use observed fluxes when doing source plane optimisation (but see further discussion of this in the \emph{Supplementary Note: Host Image Morphology Comparison} below).  

Supplementary Table~\ref{tab:mu_comparison} shows the magnification values predicted by our lens model E in comparison to the values reported in N18. 
Columns 2 and 3 give magnifications for the location of each SN image, while columns 4 and 5 are for the host galaxy images.  Column 2 repeats the SN magnifications given in the main text and Supplementary Table~\ref{tab:massmodel}, which are the expectation values from the $\mu$ distributions generated by the \lenstool MCMC sampling over model parameter space.  Column 3 gives the ``optimized'' SN magnification, which is the $\mu$ value returned by the single model instance that has the minimum $\chi^2$ value (note that this may be significantly different from the peak of the $\mu$ distribution, as for image 2 in particular).  Column 4 reports the minimum-$\chi^2$ magnification at the peak of the surface brightness profile for each galaxy image.  Column 5 reports the
ratio of the total flux to source flux of the host galaxy, which is effectively a flux-weighted harmonic mean of the magnification factors across each galaxy image. This last value is the most appropriate for comparison to the values of N18 (given in Column 6), which are computed in a similar way.  The uncertainties from N18 reflect an estimate of systematic uncertainties, derived from lens modeling variants. 

From the last two columns of Supplementary Table~\ref{tab:mu_comparison} we see that our preferred lens model E magnifications are within 1$\sigma$ of the N18 model, though our predictions are systematically lower by about 20\%.  This agrees with the assessment of systematic uncertainties from our lens model variants discussed above, which are also shown as the asymmetric error bars on the SN magnitudes in Fig.~2 in the main text, and in Extended Data Fig.~2.  As seen in those figures, a larger magnification value would make the SN {\it more} consistent with the expected luminosity of a Type Ia SN at this redshift---though it would also make it more consistent with some CC SN light curves, likely making the classification somewhat more ambiguous. 

It is important to note that the all of the model variants explored here adopted 
dPIE  distributions as the density profile for all lensing components.   It is well-documented that the choice of the density profile can strongly affect the inferred magnifications and time delays in a strong lensing system.  This is therefore another potential source of systematic uncertainty that should be explored in future lens modeling.  The models of N18 explored models using gNFW profiles, but of course did not make explicit predictions for the magnifications or time delays at the SN locations.
If alternate density profiles result in significant changes to the model-predicted magnfications for the SN images, that could in principle change the conclusions about SN classification and age constraints described here.

We expect that new lens modeling of this cluster with alternate software and different choices of constraints will also be informative, and may improve on our model predictions.  
Both the N18 modeling and the construction of lens model E and our model variants are constructed blind (without knowledge of the SN magnitudes).
Future lens modeling could incorporate the measured magnitudes as constraints, potentially yielding a more robust prediction of the time delay for the fourth image. We hope that the discovery of \SNABC  will encourage such efforts.

\subsection*{Supplementary Note: Host Image Morphology Comparison}

\begin{figure}
    \centering
    \includegraphics[width=0.75\textwidth]{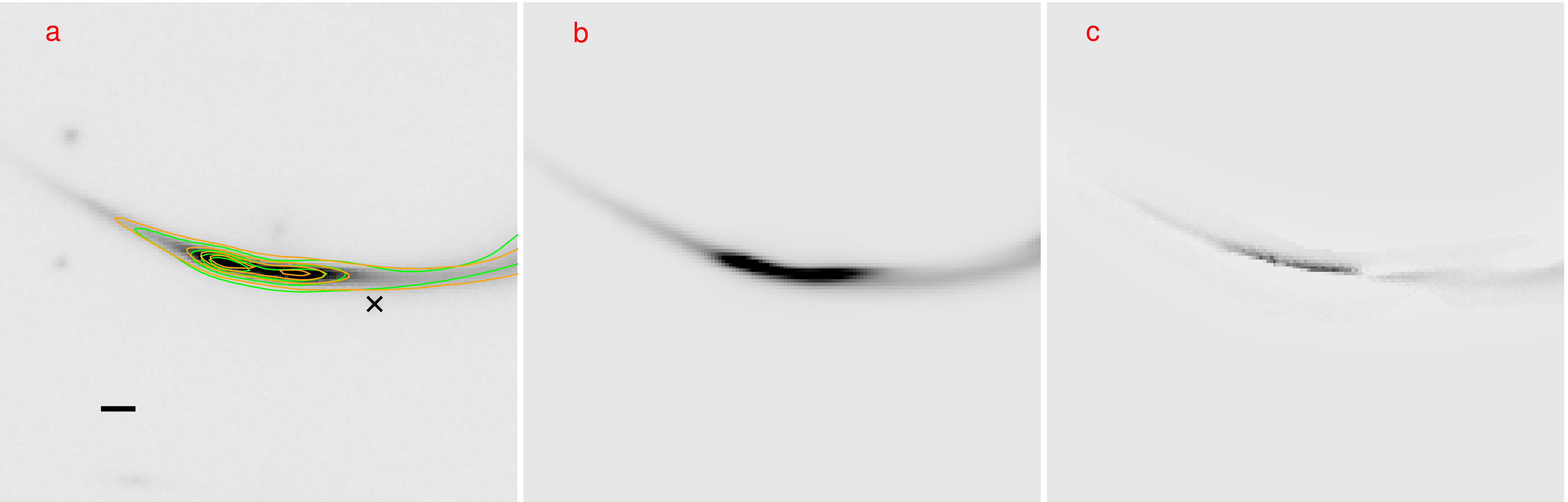}\\
    \includegraphics[width=0.75\textwidth]{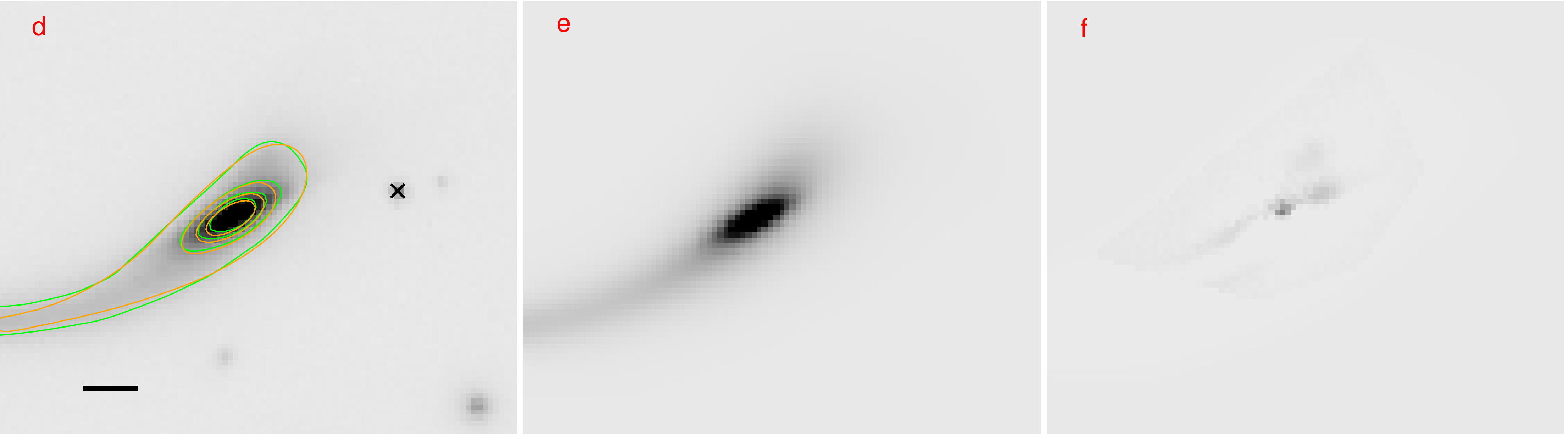}\\
    \includegraphics[width=0.75\textwidth]{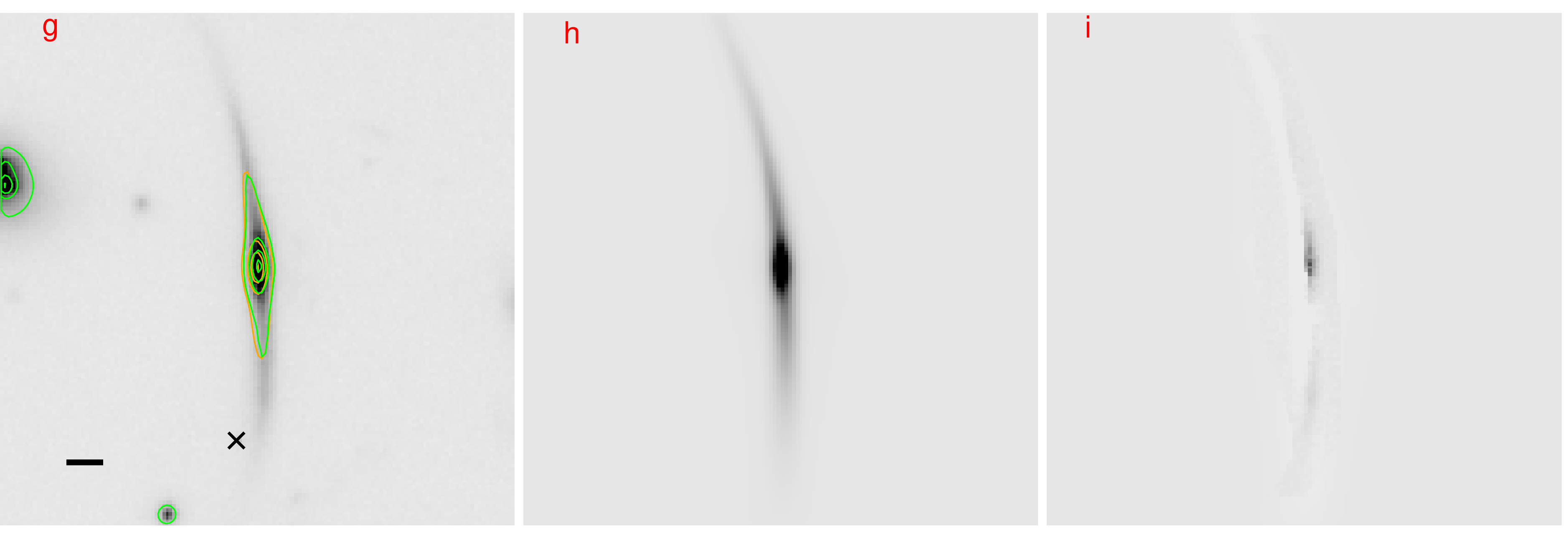}
    \caption{\textbf{Comparison of the SN host galaxy against a simulation using lens model E.} North is up and East is left. Image 1.1 is shown in the top row, 1.2 in the middle, and 1.3 in the bottom. The left column (panels a,d,g) shows each observed image in the F160W bandpass, with a scale bar indicating 1 arcsecond. The black cross highlights the location of the SN image in each case.  The central column (b,e,h) shows a simulated galaxy profile generated using lens model E, with a sum of two Sersic  profiles to represent the background galaxy. The last column (c,f,i) shows the residuals (observed-model). 
    Overlaid contours in the first column show the profile of the observed data (green) and the model (orange). The contours are logarithmically spaced and correspond to levels of 1, 2, 4, and 8$\sigma$ in detection, where $\sigma$ is the pixel to pixel background level.
    Some residual flux is apparent in the difference images, but the overall shape of each image is well matched by the simulated profile and the distortions introduced by lens model E.
    }
    \label{fig:host_model_residuals}
\end{figure}

\begin{figure}
    \centering
    \includegraphics[width=0.75\textwidth]{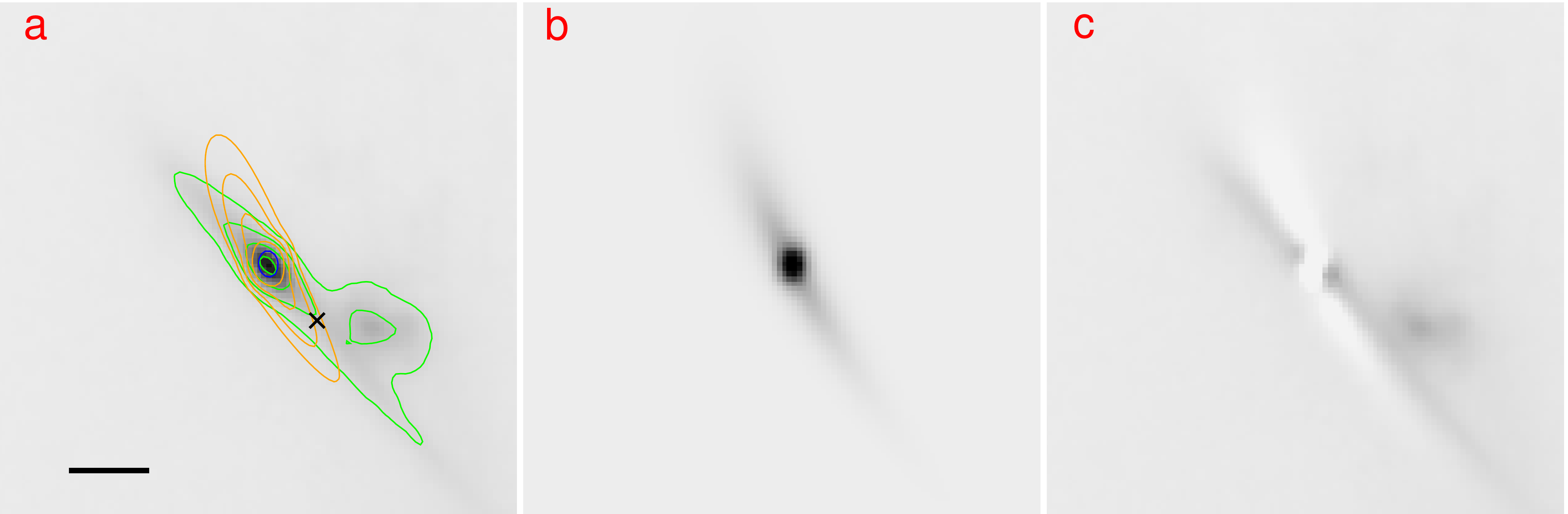}\\
     \caption{\textbf{Comparison of the SN host galaxy image 4 against a simulation using lens model E.} Same as Figure \ref{fig:host_model_residuals}, but showing galaxy image H4 (1.4).}
    \label{fig:host_image4_residuals}
\end{figure}

As a qualitative check to evaluate the accuracy of lens model E, we also simulated the overall shape of the SN host galaxy using the sum of two Sersic profiles, and then propagated this through the lens model to make predictions for the morphology of the host arcs.  The results are shown in Supplementary Fig.~\ref{fig:host_model_residuals} for host images H1-H3 (1.1-1.3).  Here we see the model produces a very good match to the observed arcs.  This is especially encouraging because the surface brightness distributions of the arcs were not used as inputs for the lens modeling. 

The predicted arc morphology for host galaxy image H4 (1.4) is shown in Supplementary Fig.~\ref{fig:host_image4_residuals}.  The residuals are more significant, showing clear asymmetric structure, which may indicate a  mismatch of rotation and/or shear angle at the location of this image.  Nevertheless, the global morphology of the arc is similar in location, size and elongation, which again could be taken as an encouraging indicator of the validity of magnification and time delay estimates for image SN4.   
For image 5 (not shown), lens model E predicts a very compact source at the location of the BCG, which is at odds with the clearly elongated radial arc of image 5 that can be seen in the HST imaging.  This may well be due to the fact that the lens model overestimates the velocity dispersion of the galaxy (see above, and Supplementary Note: Future Lens Modeling).  Reconstruction of image H5 is not as important, because the highly demagnified final transient image SN5 is not expected to be observable anyway.  Nonetheless, a more accurate reconstruction of host galaxy image H4 (and to a lesser extent image H5), should be an important metric for future lens modeling.

\subsection*{Supplementary Note: Future Work}

\subsubsection*{Future Lens Modeling}

The uncertainty in the predicted time delay from the lens model presented here is $\pm$540 days (in the observer frame).  Are there improvements to the lens modeling that could tighten this prediction?

Our lens modeling did not make use of the velocity dispersion of the BCG, measured as 390$\pm$10 km s$^{-1}$ from our MUSE data.  We did not have this measurement available prior to unblinding, and thus we have restricted ourselves to only consider fully blind lens models in this paper. However, we note that this measured value is very discrepant with the best-fit value of 700 km s$^{-1}$ from our preferred lens model, variant E (see Supplementary Table~\ref{tab:massmodel}).  Future analyses could incorporate the BCG velocity dispersion and here we speculate about the impact this would have. 

With a preliminary extension of model E we find that the time delay predictions would likely shift by $<$15 days, and the predicted magnifications for SN images 1-3 would likely be larger by as much as $\sim$2.5$\times$.  This is within the range of our systematic uncertainty estimates based on lens models A-E.  We therefore expect that inclusion of the BCG velocity dispersion will not significantly impact the SN classification or time delay conclusions, but may improve the accuracy and precision of the time delay estimate for image SN4.  

One may anticipate that future observations of the lensing cluster (e.g. with JWST) will provide more lens modeling constraints, potentially including the discovery of new multiply-imaged systems and new redshift measurements. These would substantially improve the lens model time delay predictions. Application of different lens modeling approaches could also provide some added confidence that there are not systematic biases in the lens model time delay prediction.

\subsubsection*{Future Observations}

Though the $\pm540$ day time delay uncertainty is small relative to the baseline of $>7000$ days, it nevertheless represents a long period over which the field would need to be monitored for the reappearance.  Is it feasible to expect future observations to catch the reappearance of \SNABC close to peak brightness? 
Even if the time delay uncertainty remains on the order of $\pm1$ year, it would still be reasonable to execute a follow up campaign with a cadence of approximately 2 months. Since the SN is at a redshift of $z=1.95$, a span of, say, 50 days in the observer frame is 17 days in the rest frame, comparable to the rise-time of a Type Ia SN.  Thus, it is reasonable to expect that a relatively inexpensive monitoring campaign would be able to catch the return appearance of \SNABC at or before peak brightness, ensuring a well-sampled light curve for the final SN image.

If future follow-up observations are successful in capturing the full light curve of the final fourth image, then future lens modeling could also incorporate measured SN magnifications as constraints.  Measured magnifications from lensed Type Ia SNe have been used to test lens models at both the cluster and galaxy scale \cite{nordin_lensed_2014, rodney_illuminating_2015, dhawan_magnification_2020}.  The addition of astrometric constraints for the fourth image could also significantly improve the time delay predictions for a lens model \cite{birrer_astrometric_2019}---and this could be done even with a fully blinded analysis.  Measurement of the magnification for a lensed Type Ia SN can be done without adopting strong priors from a cosmological model \cite{patel_three_2014}, meaning that one can avoid a circular constraint when the \SNABC time delay is then used for cosmology.

\subsubsection*{Future Discoveries}

In addition to follow-up observations of \SNABC, we may also hope for more discoveries of similar cluster-lensed SNe with long time delays.
A primary motivation for pursuing such events is that they can be a relatively low-cost tool for time delay cosmography.  As \SNABC shows, when the time delay is longer than a few years, the time delay measurement can be anchored at either end by just a few epochs of imaging.  If similar events are detected while the SN is still observable, one could collect a well-sampled light curve for an early and bright image using ground-based telescopes.  After waiting through the decade-long delay, the SN's reappearance can be captured with a relatively low-cost monitoring campaign. A full light curve of the final image would not be needed.  For example, with \SNABC even if the time delay is measured to only $\pm$150 days, that would be a 2\% time-delay measurement, meaning it is lens-model-limited for cosmological inferences.

The expected rate for such events is still highly uncertain, and published rate estimates to date can only be taken as extreme lower limits for the expected yield from future sky surveys \cite{riehm_near-ir_2011,li_rates_2012,petrushevska_searching_2018,petrushevska_strongly_2020}.  All of these past analyses have been 
limited to $\leq 5$ well-studied galaxy clusters.  Furthermore, they have only examined the set of {\it already known} multiply-imaged galaxies, and have 
explicitly predicted only the rate of events that would have a time delay of $<$5 years.  With these caveats, the predicted lower limits are of order 1 SN detection per year per cluster, for a deep survey with a detection limit of 27 AB mag \cite{li_rates_2012}.  At the $5\sigma$ limits of the Rubin Observatory ($i\sim23.4$ AB mag), the lower limit on that rate is reduced by about a factor of ten \cite{petrushevska_strongly_2020}. 

It is treacherous to extend these estimates to the larger population of all galaxy clusters that will be regularly observed by future wide-field surveys.  Nevertheless, let us make a crude extrapolation to motivate future work.  Consider the 1-year 2000 deg$^2$ High Latitude Survey (HLS) from the Roman Space Telescope \cite{spergel_wide_2015, troxel_synthetic_2021}, and let us conservatively apply the rate of 
$\sim$1~SN yr$^{-1}$ cluster$^{-1}$ to only the $\sim$10 most massive clusters in the HLS area. This still predicts at least 10 cluster-lensed SN detections, which is comparable to the few dozen galaxy-lensed SNe expected from the Roman SN cosmology survey \cite{pierel_projected_2021}.  
Similarly, if we apply the 10$\times$ lower rate for the Rubin Observatory to the most massive clusters in the LSST survey area, we would anticipate at least $\sim$10 detections over the 10-year survey.  This discovery rate from wide-field surveys could be enhanced with dedicated ground-based cluster surveys. \cite{stanishev_near-ir_2009,goobar_near-ir_2009,riehm_near-ir_2011}

We hope that the discovery of \SNABC will motivate an improvement over this very rough estimation of future rates.   This would require a more complete census of lensing clusters, along with lens models to predict magnifications and time delays, and measurements of star formation and stellar mass in lensed galaxies to predict the SN explosion rates.   


\bibliographystyle{naturemag}

\end{document}